\newif\ifdraft
\draftfalse

\ifdraft
\documentclass[12pt, journal, a4paper, comsoc, draftcls, onecolumn]{IEEEtran}
\else
\documentclass[journal, a4paper, comsoc]{IEEEtran}
\fi

\ifdraft
\fi

\usepackage[utf8]{luainputenc}
\usepackage[OT1]{fontenc}
\usepackage[english]{babel}

\usepackage[noadjust]{cite}

\usepackage{graphicx}
\graphicspath{{./illustrationer/}}

\usepackage[cmex10]{amsmath}
\usepackage{array}

\usepackage{booktabs}

\ifCLASSOPTIONcompsoc
  \usepackage[caption=false,font=normalsize,labelfont=sf,textfont=sf]{subfig}
\else
  \usepackage[caption=false,font=footnotesize]{subfig}
\fi

\usepackage{url}

\usepackage{microtype}

\usepackage{amssymb,mathtools}

\usepackage{minamakron}

\usepackage{pgfplots}

\pgfplotsset{
	width=256pt,
	xlabel near ticks,
	ylabel near ticks,
	label style={font=\footnotesize},
	legend style={font=\footnotesize},
	label style={font=\footnotesize},
}
\newlength{\flatFiguresHeight}
\usepackage{siunitx}

\usepgfplotslibrary{external}
{low_noise_amplifiers}
\tikzsetexternalprefix{ext_figures/}
\tikzset{external/force remake=false}

\begin{document}
	\title{Impact of Spatial Filtering on Distortion from Low-Noise Amplifiers in Massive MIMO Base~Stations}
	
	\author{Christopher~Mollén,
		Ulf Gustavsson,
		Thomas Eriksson,
		and~Erik~G.~Larsson%
		\thanks{C.~Mollén was with the Department
			of Electrical Engineering, Linköping University, 581\,83 Linköping, Sweden, when this work was done.}%
		\thanks{E.~G. Larsson is with the Department
			of Electrical Engineering, Linköping University, 581\,83 Linköping, Sweden; e-mail: erik.g.larsson@liu.se.}%
		\thanks{U.~Gustavsson is with Ericsson Research, Gothenburg, Sweden; e-mail: ulf.gustavsson@ericsson.com.}%
		\thanks{T.~Eriksson is with the Department of Electrical Engineering, Chalmers University of Technology, 412\,96 Gothenburg, Sweden; e-mail: thomase@chalmers.se.}%
		\thanks{Parts of this work were  presented at the Asilomar Conference of Signals, Systems, and Computers 2017 \cite{mollen2017analysis}.}
		\thanks{This research has been carried out in the GigaHertz centre in a joint research project financed by VINNOVA, Chalmers, Ericsson, RUAG and SAAB.  This work was also supported by Vetenskapsrådet (the Swedish research council) and ELLIIT.}}
	
	\maketitle
	
	\ifdraft\vspace{-8ex}\fi
	
	\begin{abstract}
		In massive \MIMO base stations, power consumption and cost of the low-noise amplifiers (\LNAs) can be substantial because of the many antennas.  We investigate the feasibility of inexpensive, power efficient \LNAs, which inherently are less linear.  A polynomial model is used to characterize the nonlinear \LNAs and to derive the second-order statistics and spatial correlation of the distortion.  We show that, with spatial matched filtering
		(maximum-ratio combining) at the receiver, some distortion terms combine coherently, and that the \SINR of the symbol estimates therefore is limited by the linearity of the \LNAs.  Furthermore, it is studied how the power from a blocker in the adjacent frequency band leaks into the main band and creates distortion.  The distortion term that scales cubically with the power received from the blocker has a spatial correlation that can be filtered out by spatial processing and only the coherent term that scales quadratically with the power remains.  When the blocker is in free-space line-of-sight and the \LNAs are identical, this quadratic term has the same spatial direction as the desired signal, and hence cannot be removed by linear receiver processing.  
	\end{abstract}
	
	\ifdraft\vspace{-2ex}\fi
	
	
	\IEEEpeerreviewmaketitle
	
	\section{Introduction}
	\IEEEPARstart{L}{ow-noise} amplifiers (\LNAs), which are used to amplify the weak received signal before further signal processing, are often assumed to be linear in the analysis of massive \MIMO.  Furthermore, the \LNAs often stand for a significant part of the power consumption in the receiver \cite{desset2012flexible}.  Because of the great number of radio chains that are needed to build a massive \MIMO base station, the power consumption of the hardware becomes an issue as the number of antennas is increased \cite{buzzi2016survey}.  
	For example, the total power consumption of the many \LNAs in the receiver grows large, because their power consumption cannot be decreased with an increased number of antennas, in contrast to the power amplifiers of the transmitter.
	To improve power efficiency of the \LNA, its operating point can be moved closer towards the saturation point.  
	Under such operation, the nonlinear effects of the \LNAs become more pronounced and the commonly made assumption on linearity becomes inaccurate.  
	Importantly, when a blocker (a strong undesired received signal) is received, forcing the \LNA into a nonlinear mode of operation, significant distortion can be created, which degrades the performance of the system.  
	Often the tolerance to blockers is the main factor that determines the linearity requirement of the hardware in the uplink.  
	In this paper, we analyze how the distortion from nonlinear \LNAs after spatial filtering, i.e.\ linear receiver combining, affects the performance of a massive \MIMO base station in the presence of a blocker.  
	
An important, specific question is to what extent the distortion
arising from nonlinearities is spatially correlated among the antennas
or not, and if it is, whether this correlation
matters.\footnote{Correlation of the \emph{distortion} must not be
confused with correlation of the \emph{fading} (for an in-depth
treatment of the latter, see, e.g., \cite{bjornson2017massiveBook}).
We say that the additive distortions $d_m$ and $d_{m'}$ at two
antennas $m$ and $m'$ are correlated for a given fading state if the
correlation coefficient between $d_m$ and $d_{m'}$ is non-zero,
conditioned on that fading state.  Note that in some cases, for
example if there is no randomness in the signal, then the distortion
correlation matrix will have rank one.}
Spatially uncorrelated approximations to the distortion
have been suggested and used in a range of previous
work \cite{papazafeiropoulos2018impact, ding2018massive,
zhang2018spectral,bjornson2013massive} and may be traced back to, at
least, Chapter 6.3.2 in \cite{schenk2008rf}. This model cannot,
however, be justified from basic physical principles.  Moreover, as we
show in this paper, especially in the presence of blockers, it fails
to accurately describe the characteristics of the distortion.
Notwithstanding, it is known that in certain cases (see Section IX for
details), a spatially-uncorrelated distortion model does constitute a
reasonable approximation as far as in-band error-vector magnitudes are
concerned \cite{UGUSGC14}.
	
	The correlation of the distortion in antenna arrays has also been noted in previous work.  
	For example, while using an uncorrelated model for its main analysis,   \cite{bjornson2013massive} discusses and   anticipates
	the fact that the distortion may be correlated in practice.  Reference \cite{moghadam2012correlation} uses a third-degree polynomial model to show that distortion from a multi-antenna transmitter is spatially correlated, and draws the conclusion that the correlation is negligible whenever there are two or more beamforming directions.  In \cite{suzuki2008transmitter}, a \MIMO-\OFDM system is studied and it is concluded that the performance predictions made by assuming uncorrelated receiver noise do not align with measured data.  
	The transmission from uniform linear phased arrays and free-space line-of-sight beamforming was also considered in \cite{sandrin1973spatial, hemmi2002pattern}, where the directions of the radiated distortion caused by nonlinearities were derived.
	
	Consistent with \cite{ghannouchi2009behavioral}, in this paper 
	we use behavioral models to describe the nonlinearities of the \LNAs.  
	Furthermore, the system is described
	 in continuous time.  Hence, the spatial correlation of the distortion is accurately captured.  We then analyze the
	  third-degree distortion term more closely, to draw qualitative conclusions.  The restriction to the third degree is common practice
	in the literature, see, e.g., \cite{vuolevi2003distortion}, as this commonly is the dominant term.  Additionally, in the specific  case of \LNAs, the 
	   nonlinearities are normally rather mild at foreseen operating points; thus  the third-degree term typically
	    dominates.  Our in-depth analysis of the third-degree distortion terms reveals qualitatively, and quantitatively, the impact of the nonlinearity.  
			
	The effect of nonlinear \LNAs in massive \MIMO has not yet been analyzed with a precise hardware model.  Notably, these effects cannot be analyzed with the linear, symbol-sampled models that are predominantly used in the massive \MIMO literature, 
	nor with the methodology used in \cite{papazafeiropoulos2018impact, ding2018massive, zhang2018spectral, bjornson2013massive}.

\subsection{Specific Technical Contributions}	
	
	We study a conventional receiver architecture, where each antenna is equipped with a \LNA.  A polynomial model is used to describe the nonlinear effects of the \LNAs and the Itô-Hermite polynomials are employed to derive the autocorrelation of the additional error term of the symbol estimate that is caused by the nonlinear \LNAs.  We focus on the analog front-end, within which the \LNAs often constitute the main contributors to distortion from nonlinearities.  The presence of additional nonlinearities, for example in mixed-signal devices, \ADCs notably, would aggravate the problems with blocking. (We do not model the \ADCs here, since the polynomial model is a relatively poor model due to the discontinuous nature of quantization.)
	
	The special case of free-space line-of-sight, where each signal travels on a single path straight from the transmitter to the base station, is then studied.
	This gives important insights into the basic phenomenology that appears when nonlinearities are present in the receiver hardware.  Since strong blockers often appear in line-of-sight, this is a highly relevant, practical case.  Line-of-sight is also a common propagation condition, especially in the mmWave band.  Also it is likely that many laboratory
	 experiments, conformance tests, and tests that deal with sensitivity to blockers will primarily, or at least
	initially, be conducted in anechoic chambers with conditions close to free-space line-of-sight propagation.  An in-depth understanding of this case is therefore imperative.
	
	We show that the distortion combines coherently when using maximum-ratio combining, both in the presence and absence of a blocker, and that the received \SINR is limited by the linearity of the \LNAs.  The same is true for any linear decoder  that does not take the
	 spatial correlation of the distortion into account.  
	In the case of a blocker in free-space line-of-sight, it is shown that the nonlinear distortion gives rise to two
	kinds of error terms: one that scales quadratically with the received power from the blocker and one that scales cubically.  
	With sufficiently
	   many antennas, spatial processing can suppress the cubic term.  However, the quadratic
	term combines in the same way as the desired signal in the decoding.
	
\subsection{Other Related Work}
	
		Recently, massive \MIMO base stations with another kind of receiver nonlinearity---low-resolution \ADCs---have received
		 some attention \cite{verenzuela2016hardware,sarajlic2017low,jacobsson2017throughput}.  In \cite{mollen2016uplink, mollen2017achievable}, it was
		  shown that the quantization distortion from low-resolution \ADCs combines noncoherently when the channel has a high degree of frequency selectivity and the signal that is to be decoded is received with a small power compared to
		   the interference and noise.  However, in scenarios with frequency-flat channels and
	a single received signal with high \SNR, the quantization distortion combines in the same way as the desired signal.  
	This is in
		     line with the findings in this paper, which is natural since both an \ADC and an \LNA have finite
		      dynamic ranges that lead to signal clipping.

	Finally, while our paper deals with the uplink, some comments on the downlink are in order.
	In the downlink, it is known from analytical calculations \cite{mollen2017nonlinear2ArXiv,mollen2016outofbandArXiv} and simulations \cite{UGUSGC14} that power amplifier nonlinearities cause distortion that is correlated among the antennas and, thus, adds up constructively in specific spatial directions.  These spatial directions depend on the beamforming weights which, in turn, depend on the channel responses of the terminals targeted by the beamforming.  This holds both for in-band distortion and out-of-band radiation.  
	Some studies of the effect of nonlinear amplification in the massive \MIMO downlink employ a symbol-sampled discrete-time signal model \cite{blandino2017analysis,zou2015impact}.  Such symbol-sampled signal models cannot accurately describe the distortion effects of a nonlinearity in a continuous-time communication system, especially not phenomena that arise out-of-band.

	The work in \cite{zhao2017energy,yu2017energy} proposed a hardware architecture for the receiver in
	 a massive \MIMO base station that only employs a single \LNA for the whole array.  That work, however, used a simplistic hardware model that did not take nonlinear effects into account.  

	\section{Preliminaries: Nonlinearities and Passband Signals}\label{sec:nonlinear_model}
	In this section, we give a description and self-contained derivation of some mathematical results that will be exploited later in
	the paper.
	
	We consider a real-valued static nonlinearity $\hat{\symcal{A}}$ acting on a real-valued passband signal, i.e.\ a signal $\hat{x}(t)$ whose Fourier transform
	\begin{align}
		\symsfit{\hat{x}}(f) = \int_{-\infty}^{\infty} \hat{x}(t) e^{-2\pi t f} \symrm{d}t
	\end{align}
	is zero outside a band of width $B < f_\text{c}$ centered around $f_\text{c}$:
	\begin{align}
		\symsfit{\hat{x}}(f) = 0, \quad \text{when } |f| \notin [f_\text{c} - B/2, f_\text{c} + B/2].
	\end{align}
	The most general nonlinearity with memory can be described by a Volterra series of degree $\Pi$ \cite{schetzen1980volterra}:
	\begin{align}\label{eq:988907232}
		\hat{\symcal{A}}(\hat{x}(t)) = \hat{b}_0 + \sum_{\varpi=1}^{\Pi} \int_{-\infty}^{\infty} \mathclap{\cdots}\,\, \int_{-\infty}^{\infty}\! \hat{b}_\varpi(\tau_1, \ldots, \tau_\varpi) \prod_{\varpi'=1}^{\varpi} \hat{x}(t - \tau_{\varpi'}) \symrm{d}\tau_{\varpi'}.
	\end{align}
	Depending on the type of memory that the nonlinearity has, this model can be simplified and rephrased in terms of the equivalent baseband representations of the in- and outputs.  Some common memory types are given in Table~\ref{tab:deg_memorylessness}.  This paper focuses on memoryless and quasi-memoryless nonlinearities.

	\begin{table*}
	\ifdraft\renewcommand{\baselinestretch}{1}\fi
	\centering
	\caption{Degrees of memorylessness}
	\label{tab:deg_memorylessness}
	\ifdraft\smallskip\fi
	\rule{0pt}{0pt}\clap{\begin{tabular}{>{\raggedright\ifdraft\setlength{\baselineskip}{.6\baselineskip}\fi}p{8em}>{\raggedleft}p{12.5em}>{\raggedright\ifdraft\setlength{\baselineskip}{.6\baselineskip}\fi}p{12em}>{\raggedright\ifdraft\setlength{\baselineskip}{.6\baselineskip}\fi\arraybackslash}p{13.5em}}
			\toprule
			nonlinearity type & \raggedright kernel $\hat{b}_n(\tau_1, \ldots, \tau_n) = $ & baseband description & remark \\
			\midrule
			with memory & $\hat{b}_n(\tau_1, \ldots, \tau_n)$ & pure Volterra& used for the most general systems \\
			\addlinespace
			with two-dimensional memory & $\displaystyle\hat{b}_n(\tau_1, \tau_2 - \tau_1) \prod_{n'=3}^{n}\delta(\tau_{n'} - \tau_2)$ & generalized memory polynomial & used when the mass of the kernels outside a two-dimensional slice through the diagonal of the space $(\tau_1, \ldots, \tau_n)$ can be neglected\\
			\addlinespace
			with one-dimensional memory & $\displaystyle\hat{b}_n(\tau_1) \prod_{n'=2}^{n} \delta(\tau_{n'}-\tau_1)$ & memory polynomial & used when the mass of the kernels outside the diagonal can be neglected\\
			\addlinespace
			quasi-memoryless & $\displaystyle\hat{b}_n \prod_{n'=1}^{n} \delta(\tau_{n'})$, $\hat{b}_n\in\symbb{C}$ & polynomial model with odd-degree terms and complex coefficients & used when the kernels are approximately constant in the frequency band of the passband output signal\\
			\addlinespace
			memoryless & $\displaystyle\hat{b}_n \prod_{n'=1}^{n} \delta(\tau_{n'})$, $\hat{b}_n\in\symbb{R}$ & polynomial model with odd-degree terms and real coefficients & used when the kernels are perfectly frequency flat \\
			\bottomrule
		\end{tabular}}\rule{0pt}{0pt}
	\end{table*}

	A system is said to be \emph{quasi-memoryless} when the multi-dimensional Fourier transform 
	\begin{align}
		\int_{-\infty}^{\infty}\cdots\int_{-\infty}^{\infty} \hat{b}_\varpi(\tau_1, \ldots, \tau_\varpi) \prod_{\varpi'=1}^{n} e^{-j2\pi f_{\varpi'}\tau_{\varpi'}} \symrm{d}\tau_{\varpi'}
	\end{align}
	of the kernel $\hat{b}_\varpi(\tau_1, \ldots, \tau_\varpi) = b_\varpi$ is constant over the band around the center frequency $(f_1,\ldots,f_{\varpi}) = (f_\text{c}, \ldots, f_\text{c})$ of the passband signal $\hat{x}(t)$ for all $\varpi$ \cite{raich2002modeling}.  If we only are interested in the action of the nonlinearity on the frequency component around the center frequency, then the kernels can be simplified, as shown in Table~\ref{tab:deg_memorylessness}, to the complex gains $\{\hat{b}_\varpi\}$ of the Fourier transforms at $(f_1,\ldots,f_\varpi) = (f_\text{c}, \ldots, f_\text{c})$.  
	
	In case of a quasi-memoryless nonlinearity, the output signal in \eqref{eq:988907232} thus simplifies into:
	\begin{align}\label{eq:real_pb_nonlinearity}
		\hat{y}(t) = \hat{\symcal{A}}(\hat{x}(t)) = \sum_{\varpi=0}^{\Pi} \hat{b}_\varpi \hat{x}^\varpi(t).
	\end{align}
Conceptually, this polynomial approximation may also be justified through the  Weierstrass approximation theorem that states that a continuous function can be approximated arbitrarily well by a polynomial on a closed interval.

	The output of the real nonlinearity acting on the passband signal $\hat{x}(t)$ in \eqref{eq:real_pb_nonlinearity} consists of a sum of spectral components that are distinct in the frequency domain if $\Pi B < f_\text{c}$.  Each spectral component is concentrated around a multiple of the frequency $f_\text{c}$, see \cite[Fig.~5.3]{middleton1996introduction}.  Denote the baseband representation of the spectral component around the frequency $f_\text{c}$ by
\begin{align}\label{eq:baseband_signal}
y(t) = \symcal{B}\left(\hat{y}(t) e^{-j2\pi f_\text{c}t}\right),
\end{align}
where $\symcal{B}$ is an ideal lowpass filter with cut-off frequency $f_\text{c}/2$.  The spectral component described by $y(t)$ can be given in terms of the baseband equivalent of the input signal
\begin{align}
x(t) = \symcal{B}\left(\hat{x}(t) e^{-j2\pi f_\text{c}t}\right)
\end{align}
and the baseband equivalent $\symcal{A}$ of the nonlinearity:
\begin{align}\label{eq:polynomial_model}
y(t) = \symcal{A}(x(t)) = \sum_{\varpi \leq \Pi:\text{odd}} b_\varpi x(t) |x(t)|^{\varpi-1},
\end{align}
where the sum is over all odd indices $\varpi \leq \Pi$, for some coefficients $b_\varpi$, $\varpi=1,3,5,\ldots, \Pi$ that are equal to the passband coefficients up to a scaling:
\begin{align}\label{eq:91288838381}
b_\varpi = {\varpi \choose \frac{\varpi+1}{2}} \hat{b}_\varpi.
\end{align}
For completeness, the derivation of \eqref{eq:polynomial_model} from \eqref{eq:real_pb_nonlinearity} is given in Appendix~\ref{app:polynomial_model}.  The baseband equivalent in \eqref{eq:polynomial_model} of the nonlinearity in \eqref{eq:real_pb_nonlinearity} is called its \textit{polynomial model} \cite{schreurs2009rf, raich2002modeling}.  

The complex Itô generalization of the Hermite polynomials \cite{ito1952complex, dunkl2014orthogonal},
\begin{align}\label{eq:complex_hermite_polynomial}
H_{\varpi,\varpi'}(x,x^*) \defas \varpi!\varpi'! \sum_{i=0}^{\min\{\varpi,\varpi'\}} \frac{(-1)^i}{i!} \frac{x^{\varpi-i} (x^*)^{\varpi'-i}}{(\varpi-i)!(\varpi'-i)!},
\end{align}
has the following property:
\begin{multline}\label{eq:910661188823}
\Exp\left[H_{\varpi,\chi}(X,X^*) H^*_{\varpi',\chi'}(Y,Y^*) \right]\\
= \varpi!\chi! \delta[\varpi-\varpi'] \delta[\chi-\chi'] \Exp[XY^*]^{\varpi} \Exp[X^*Y]^{\chi},
\end{multline}
where $X$ and $Y\sim\CN(0,1)$ are standard circularly symmetric complex jointly Gaussian variables, $p_{X,Y}(x,y)$, $p_X(x)$ and $p_Y(y)$ are the joint probability density function of the variables $X$ and $Y$, the density of $X$ and the density of $Y$ respectively.  Equation \eqref{eq:910661188823} can be shown by using the orthogonality of the polynomials and the complex generalization of Mehler's formula \cite[ref.\ to as “Poisson kernel”]{ismail2016analytic}:
\begin{multline}
\frac{p_{X,Y}(x,y)}{p_{X}(x)p_Y(y)}\\
= \sum_{\varpi=0}^{\infty} \sum_{\varpi'=0}^{\infty} \frac{\Exp[XY^*]^\varpi \Exp[X^*Y]^{\varpi'}}{\varpi!\varpi'!} H_{\varpi,\varpi'}(x,x^*) H^*_{\varpi,\varpi'}(y,y^*).
\end{multline}

If the signal $x(t)$ is Gaussian, has zero mean and unit power, an orthogonal basis for the space of complex polynomials of the kind in \eqref{eq:polynomial_model} is given by a subset of the complex Itô generalization of the Hermite polynomials in \eqref{eq:complex_hermite_polynomial}, namely by:
\begin{align}
H_\varpi(x) &\defas H_{\frac{\varpi+1}{2}, \frac{\varpi-1}{2}}(x,x^*)\\
&= \sum_{n=0}^{\frac{\varpi-1}{2}} (-1)^n n! {\frac{\varpi+1}{2} \choose n} {\frac{\varpi-1}{2} \choose n} x |x|^{\varpi-1-2n}, \label{eq:82893882991092}
\end{align}
for $\varpi=1,3,5,\ldots, \Pi$.  The first polynomials of this kind are given in Table~\ref{tab:complex_hermite_polynomials}.  

\begin{table}
	\ifdraft\linespread{1}\fi
	\centering
	\caption{Complex Hermite polynomials}
	\label{tab:complex_hermite_polynomials}
	\begin{tabular}{@{}r@{\,}c@{\,}l@{\hspace{-20em}}r@{}}
		&&&$x|x|^8=H_9(x) + 20 H_7(x) + 120 H_5(x) + 240 H_3(x) +120 H_1(x)$\\
		$H_1(x)$ & $=$ & $x$ & $x|x|^6=H_7(x) + 12 H_5(x) + 36 H_3(x) + 24 H_1(x)$\\
		$H_3(x)$ &$ = $&$x|x|^2 - 2x$ & $x|x|^4 = H_5(x) + 6H_3(x) + 6 H_1(x)$\\
		$H_5(x)$ &$=$& $x|x|^4 - 6x|x|^2 + 6x$ & $x|x|^2= H_3(x) + 2 H_1(x)$\\
		$H_7(x)$ & $=$ & $x|x|^6 -12 x|x|^4 + 36 x|x|^2 -24 x$ & $x=H_1(x)$\\
		$H_9(x)$ & $=$ & $x|x|^8 - 20 x|x|^6 + 120 x|x|^4 - 240 x|x|^2 + 120 x$ & \\
	\end{tabular}
\end{table}	

The baseband equivalent of the nonlinearity \eqref{eq:polynomial_model} can thus be rewritten as
\begin{align}\label{eq:881008}
y(t) = \sum_{\varpi\leq \Pi : \text{odd}} a_\varpi H_\varpi(x(t))
\end{align}
for some coefficients $\{a_\varpi\}$.  Using the property in \eqref{eq:910661188823}, the autocorrelation of the baseband signal can be obtained as:
\begin{align}
R_{yy}(\tau) &= \Exp[y(t) y^*(t-\tau)]\\
&= \sum_{\varpi\leq \Pi:\text{odd}} |a_\varpi|^2 \left(\frac{\varpi+1}{2}\right)! \left(\frac{\varpi-1}{2}\right)! \, r_x(\tau) |r_x(\tau)|^{\varpi-1}.\label{eq:86912910013}
\end{align}
Since the terms in \eqref{eq:82893882991092} are pairwisely uncorrelated and $H_1(x) = x$ in the first term, the output can be decomposed into an undistorted linear term and a distortion term:
\begin{align}\label{eq:decomposition_amp_sig}
y(t) = a_1 x(t) + d(t),
\end{align}
where the distortion is uncorrelated to the linear term and is given by
\begin{align}\label{eq:uncorrelated_distortion}
	d(t) = \sum_{\varpi=3,5,\ldots,\Pi} a_\varpi H_\varpi(x(t)).
\end{align}
The autocorrelation function of the distortion is given by:
\begin{align}
	R_{dd}(\tau) = \sum_{\varpi=3,5,\ldots,\Pi} |a_\varpi|^2 \left(\frac{\varpi+1}{2}\right)! \left(\frac{\varpi-1}{2}\right)! \, r_x(\tau) |r_x(\tau)|^{\varpi-1}.
\end{align}

Another fact that will be used is that an input cross-correlation is transformed by nonlinearities in the same way as the autocorrelation in \eqref{eq:86912910013}.  Let $x_1(t), x_2(t) \sim \CN(0,1)$ be jointly Gaussian with cross-correlation $R_{x_1x_2}(\tau) \defas \Exp[x_1(t) x^*_2(t-\tau)]$.  When these signals are input to two different nonlinear systems with outputs
\begin{align}
y_1(t) &= \symcal{A}_1(x_1(t)) = \sum_{\varpi\leq \Pi:\text{odd}} a_{1\varpi} H_\varpi(x_1(t)),\\
y_2(t) &= \symcal{A}_2(x_2(t)) = \sum_{\varpi\leq \Pi:\text{odd}} a_{2\varpi} H_\varpi(x_2(t)),
\end{align}
the cross-correlation of the outputs is given by
\begin{align}
&R_{y_1y_2}(\tau) \defas \Exp[y_1(t) y^*_2(t-\tau)]\\
&\quad= \sum_{\mathclap{\varpi\leq \Pi: \text{odd}}} a_{1\varpi} a^*_{2\varpi} \left(\frac{\varpi+1}{2}\right)! \left(\frac{\varpi-1}{2}\right)! r_{x_1x_2}(\tau)|r_{x_1x_2}(\tau)|^{\varpi-1}.\label{eq:828129859191761}
\end{align}
	
	\section{System Model}
\begin{figure}[t!]
		\centering\footnotesize
		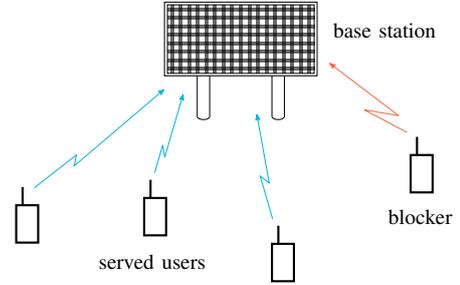
		\caption{The base station receives signals from multiple served users that are to be decoded.  Additionally, it receives an interfering signal, which could be a signal from a user that is served by another base station or a signal from a malicious transmitter.}
		\label{fig:setup1888111293}
	\end{figure}

	We analyze the uplink transmission from $K$ single-antenna users to a base station with $M$ antennas.  Additionally, an undesired transmitter 
	(blocker) is present, whose signal interferes with the received signals from the served users.  The setting is depicted in Figure~\ref{fig:setup1888111293}.

	The transmitted signals are generated by pulse amplitude modulating discrete symbols $x_k[n]$ with symbol period $T$ and the pulse-shaping filter $p(\tau)$:
	\begin{align}
		x_{k}(t) = 
		\begin{cases}
			\sum_{n=-\infty}^{\infty} \sqrt{P_k} x_k[n] p(t - nT),&\text{if } k = 1,\ldots,K,\\
			\sum_{n=-\infty}^{\infty} \sqrt{P_k} x_k[n] p(t - nT) e^{j2\pi B t},&\text{if } k = K+1,
		\end{cases}
	\end{align}
	where $k = 1, \ldots, K$ are indices of served users and $k=K+1$ is the index of the blocker.  The pulse-shaping filter is assumed to have bandwidth $B$ and to be strictly limited to the frequency band $[-B/2, B/2]$.  The power of the symbols is normalized such that $\Exp[|x_k[n]|^2] = 1$ and $\int_{-\infty}^{\infty}|p(\tau)|^2\symrm{d}\tau = T$, so that $P_k$ represents the transmit power of transmitter $k$.  The blocker uses the same pulse shape as the served users but transmits in the adjacent frequency band  with center frequency $f=B$.  The blocker could, for example, model a single-antenna user that belongs to another communication system that transmits in the right adjacent band.  

	In order to use the properties of the Itô-Hermite polynomials, the \emph{received} signals should be Gaussian.  To ensure that, it is assumed that the symbols $x_k[n]$ are circularly symmetric, complex Gaussian, which is a good model for \OFDM signals.  Often when multiple signals are multiplexed the received signals are close to Gaussian, even if the transmit symbols are not Gaussian, due to the law of large numbers.  
	
	The channel from transmitter $k \in \{1, \ldots, K+1\}$ to base station antenna $m$ is denoted by the coefficient $h_{km}\in\symbb{C}$.  Thus, the signal that is received at antenna $m$ is given by:
	\begin{align}\label{eq:88181833}
		u_m(t) = \sum_{k=1}^{K+1} h_{km} x_k(t) + z_m(t),
	\end{align}
	where $z_m(t)$ is a stationary Gaussian process that is used to model the thermal noise of the receiver hardware.  It will be assumed that the noise is independent across the antennas $m$ and is white over   the three adjacent frequency bands.  Specifically, it has constant power spectral density $N_0$ over the band $[-3B/2, 3B/2]$ and power spectral density equal to zero outside this bandwidth.  The channel is assumed to be frequency flat in order to ease the notation.  
	
\begin{figure}
		\centering
		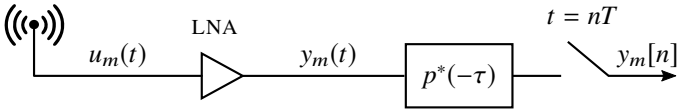
		
		\caption{A schematic view of the studied receiver model.}
		\label{fig:rx_model929293}
	\end{figure}

	The received signals are then filtered by a receive filter prior to sampling, as shown in Figure~\ref{fig:rx_model929293}.  In practice, the filtering is done in the digital domain in an intermediate, oversampled stage.  Mathematically, however, such a receiver chain is equivalent to the analog one in Figure~\ref{fig:rx_model929293}, as long as the oversampling factor is large enough.  To simplify the exposition, we will therefore analyze the system in Figure~\ref{fig:rx_model929293} without an intermediate, oversampled stage.
		
	Upon reception, the weak signal is amplified by an \LNA.  The operation of the \LNA at antenna $m$ will be denoted by $\symcal{A}_m$ and the amplified received signal is given by:
	\begin{align}\label{eq:9183888}
		y_m(t) = \symcal{A}_m\left(u_m(t)\right),
	\end{align}
	Note that the amplification operation $\symcal{A}_m$ is different at different antennas $m$ in general.
	
	The \LNAs will be modeled as quasi-memoryless by a polynomial model as described in Section~\ref{sec:nonlinear_model}.  An example of coefficients for the polynomial model obtained from measurements on a GaN (Gallium Nitride) amplifier designed for operation at \SI[round-precision=2]{2.1}{GHz} is given in the same Table~\ref{tab:amp_coeffs}.  The input and output data were sampled at \SI{200}{MHz} and the signal bandwidth was \SI{40}{MHz}.  
	The coefficients have been rescaled such that the unit power input signal is \SI{9}{dB} below the saturation point, which would imply that the signal is below the saturation point 99.99\,\% of the time.  
	The corresponding Hermite coefficients are given in Table~\ref{tab:amp_coeffs} for an input signal with unit power.
	
	The fact that the received signal has an arbitrary power has to be considered when rewriting the polynomial model using the Itô-Hermite polynomials, c.f.\ \eqref{eq:881008}:
	\begin{align}\label{eq:00919922}
	\symcal{A}_m(u_m(t)) = \sum_{\varpi \leq \Pi: \text{odd}} a_{\varpi m} \sigma_{u_m}^\varpi H_\varpi\left(\frac{u_m(t)}{\sigma_{u_m}}\right),
	\end{align}
	where the complex coefficients $\{a_{\varpi m}\}$ are linear combinations of the coefficients $\{b_{\varpi m}\}$ and different powers of the signal power:
	\begin{align}
	\sigma_{u_m}^2  = \Exp\left[|u_m(t)|^2\right].	
	\end{align}
	For example, if $b_{m\varpi} = 0$ for all $\varpi > 9$, the Hermite coefficients are given by:
	\begin{align}
	a_{1m} &= b_{1m} + 2 \sigma_{u_m}^2 \! b_{3m} + 6 \sigma_{u_m}^4 \! b_{5m} {+} 24 \sigma_{u_m}^6 \! b_{7m} {+} 120 \sigma_{u_m}^8 \! b_{9m}\label{eq:53836351982736}\\
	a_{3m} &= b_{3m} + 6 \sigma_{u_m}^2 b_{5m} + 36 \sigma_{u_m}^4 b_{7m} + 240 \sigma_{u_m}^6 b_{9m}\\
	a_{5m} &= b_{5m} + 12 \sigma_{u_m}^2 b_{7m} + 120 \sigma_{u_m}^4 b_{9m}\\
	a_{7m} &= b_{7m} + 20 \sigma_{u_m}^2 b_{9m}\\
	a_{9m} &= b_{9m}.\label{eq:720520367898052}
	\end{align}
	
	\begin{table}
		\centering
		\caption{Polynomial coefficients fitted to measured data from a GaN~amplifier \cite{ericsson2016polynom_model_pa}.}
		\label{tab:amp_coeffs}
		\begin{tabular}{l@{\hspace{1em}}l}
			\toprule
			polynomial coefficients & Hermite coefficients\\
			\midrule
			$b_1=\num[round-mode=figures, round-precision=3, group-separator={}]{0.99995200}-j\num[round-mode=figures, round-precision=3, group-separator={}]{0.00981788}$&$a_1=\num[round-mode=figures, round-precision=3, group-separator={}]{0.925351833}-j\num[round-mode=figures, round-precision=3, group-separator={}]{1.93167e-03}$\\
			$b_3=-\num[round-mode=figures, round-precision=3, group-separator={}]{7.78231181e-03}+j\num[round-mode=figures, round-precision=3, group-separator={}]{0.0149617}$&$a_3=-\num[round-mode=figures, round-precision=3, group-separator={}]{0.0427976467}+j\num[round-mode=figures, round-precision=3, group-separator={}]{2.95963e-03}$\\
			$b_5=-\num[round-mode=figures, round-precision=3, group-separator={}]{2.69300297e-02}-j\num[round-mode=figures, round-precision=3, group-separator={}]{0.00736869}$&$a_5=-\num[round-mode=figures, round-precision=3, group-separator={}]{2.90982131e-03}-j\num[round-mode=figures, round-precision=3, group-separator={}]{1.19691e-03}$\\
			$b_7=\num[round-mode=figures, round-precision=3, group-separator={}]{6.54370219e-03}+j\num[round-mode=figures, round-precision=3, group-separator={}]{0.00165554}$&$a_7=-\num[round-mode=figures, round-precision=3, group-separator={}]{2.54033414e-03}-j\num[round-mode=figures, round-precision=3, group-separator={}]{6.2691e-04}$\\
			$b_9=-\num[round-mode=figures, round-precision=3, group-separator={}]{4.54201816e-04}-j\num[round-mode=figures, round-precision=3, group-separator={}]{0.00011412}$&$a_9=-\num[round-mode=figures, round-precision=3, group-separator={}]{4.54201816e-04}-j\num[round-mode=figures, round-precision=3, group-separator={}]{1.1412e-04}$\\
			\bottomrule
		\end{tabular}
	\end{table}
	
	It should be noted that $\sigma_{u_m}^2$ depends on the time $t$, because the received signal $u_m(t)$ is a cyclostationary signal.  The Hermite coefficients $\{a_{\varpi m}\}$ therefore also depend on $t$.  For tractability, however, the dependence of the coefficients on time will be neglected, by replacing $\sigma_{u_m}^2$ with the signal power $\int_{0}^{T} \sigma_{u_m}^2/T \symrm{d}t$ in \eqref{eq:53836351982736}--\eqref{eq:720520367898052}.  Many practical choices of pulses $p(\tau)$ result in signals whose energy is evenly spread in time, especially those with a small excess bandwidth.  For such pulses, it is a reasonable approximation to use constant Hermite coefficients.  To obtain expressions that are valid for any pulse, it is straightforward to avoid the approximation by taking the dependency on time into consideration in what follows.  However, the resulting expressions are less insightful as they will contain terms, in which the pulse and the coefficients are inseparable.  
	
	Because of the property of the Itô-Hermite polynomials in \eqref{eq:910661188823}, the amplifier output, as given by the expansion in \eqref{eq:00919922}, is a sum of uncorrelated signals, each defined by:
	\begin{align}
		u_{\varpi m} (t) \defas \sigma_{u_m}^\varpi H_\varpi\left(\frac{u_m(t)}{\sigma_{u_m}}\right).
	\end{align}
	Just like in \eqref{eq:decomposition_amp_sig}, we partition the amplified signal into two components:
	\begin{align}
		y_m(t) = a_{1m} u_m(t) + d_m(t),
	\end{align}
	where the uncorrelated distortion is given by:
	\begin{align}
		d_m(t) \defas \sum_{3\leq\varpi\leq\Pi:\text{odd}} a_{\varpi m} u_{\varpi m}(t).
	\end{align}
	
	In the symbol-sampled system model, the digital signal is obtained through demodulation with the matched filter $p^*(-\tau) / T$, which is scaled by the symbol period $T$ to make the variance of the sampled noise
	\begin{align}
		z_m[n] \defas \frac{1}{T} {\left(p^*(-\tau) \star z_m(\tau)\right)}(t) \Bigr|_{t=nT} \sim \CN(0,N_0/T),
	\end{align}
	equal to $N_0/T$.  The output of the matched filter is given by
	\begin{align}
		\bar{y}_m(t) = \sum_{\varpi\leq \Pi:\text{odd}} a_{\varpi m} \bar{u}_{\varpi m}(t),
	\end{align}
	where the individual terms are given by
	\begin{align}\label{eq:91828988}
		\bar{u}_{\varpi m}(t) = \frac{1}{T} \left(p^*(-\tau) \star u_{\varpi m}(\tau) \right)(t).
	\end{align}
	The matched-filter output is then sampled:
	\begin{align}
		y_m[n] &= \bar{y}_m(nT)\label{eq:0817176}\\
		&= \sum_{\varpi\leq\Pi:\text{odd}} a_{\varpi m} \bar{u}_{\varpi m}(nT).
	\end{align}
	The signal part of the first term, the linear term, is denoted $u_m[n] \defas \bar{u}_{1m}(nT)$.  The other terms, which represent the uncorrelated distortion, are denoted $d_m[n] \defas y_m[n] - a_{1m} u_m[n]$.  If we assume perfect time synchronization, i.e.\ that the sampling instants are $t=nT$ and that the pulse $p(\tau)$ is a root-Nyquist pulse of parameter $T$, the linear part of the signal can be given as:
	\begin{align}
		u_m[n] = \sum_{k=1}^{K} \sqrt{P_k} h_{km} x_k[n] + z_m[n].
	\end{align}
	It is noted that the blocker does not affect this term because its signal and the receive filter do not overlap in frequency.  The channel will be assumed to be normalized, such that:
	\begin{align}
		\Exp\left[|h_{km}|^2\right] = 1.
	\end{align}
	In this way, the power $P_k$ is the average received power from user $k$.
	
	The estimate of the transmitted symbol of user $k = 1, \ldots, K$ is obtained by decoding the digital signal:
	\begin{align}
		\hat{x}_k[n] &\defas \sum_{m=1}^{M} w_{km} y_m[n]\label{eq:8109293}\\
		&= \sum_{m=1}^{M} a_{1m} w_{km} u_m[n] + \! \sum_{m=1}^{M} w_{km} d_m[n],\label{eq:981890273765619723}
	\end{align}
	where $\{w_{km}\}$ are the weights of the linear decoder of user $k$.  The additional error in the estimate due to the nonlinear distortion is thus given by the last sum in \eqref{eq:981890273765619723}:
	\begin{align}
		e_k[n] \defas \sum_{m=1}^{M} w_{km} d_m[n].\label{eq:9777662553792}
	\end{align}
	
	Here it is assumed that there is no temporal processing.  In a perfectly linear frequency-flat channel, this would not be a limitation.  If the distortion $d_m[n]$ has an autocorrelation that is nontrivial, however, a frequency-selective decoder could suppress the distortion better. 

	\section{Effect of LNAs on Decoding}
	From the expression for the symbol estimate in \eqref{eq:981890273765619723}, it can be seen that the nonlinear \LNAs affect the symbol estimates in two ways:
	\begin{enumerate}
		\item A multiplicative distortion of the decoding weights.
		\item An additive distortion of the symbol estimates.
	\end{enumerate}
	To evaluate these two effects, we will apply the following so-called, \emph{use-and-then-forget bound} on the capacity.
	This bound is a rigorous, yet fairly simple information-theoretic technique that is  often used for
	performance analysis in massive \MIMO. 
	\begin{theorem}
		An achievable rate for the link in \eqref{eq:981890273765619723} is given by:
		\begin{align}\label{eq:897376688}
			R_k = \log(1 + \SINR_k),
		\end{align}
		where the $\SINR_k$ is called the ``effective \SINR'' and is given by:
		\begin{align}\label{eq:78919859190191}
			\SINR_k = \frac{\left|\Exp[\hat{x}_k[n] x_k^*[n]]\right|^2 / \Exp[|x_k[n]|^2]}{\Exp\left[|\hat{x}_k[n]|^2\right] - \left|\Exp[\hat{x}_k[n] x_k^*[n]]\right|^2 / \Exp[|x_k[n]|^2]}
		\end{align}
	\end{theorem}
	Detailed discussions and proofs can be found,
	for example, in  \cite{marzetta2016fundamentals, bjornson2017massiveBook}.  In the context of a nonlinear system, 
	application of the use-and-then-forget bound technique results in  the following theorem.
	\begin{corollary}
		The effective \SINR of a system with nonlinear \LNAs is
		\begin{align}\label{eq:8281882132}
			\SINR_k = \frac{P_k |g_k|^2}{\sum_{k'=1}^{K} P_{k'} \tilde{I}_{kk'} + M N_0/T + D},
		\end{align}
		where the decoding gain is given by:
		\begin{align}\label{eq:1992811123331}
			g_k \defas \sum_{m=1}^{M} \Exp\left[a_{1m}{w}_{km} h_{km}\right],
		\end{align}
		the interference from user $k'$ to user $k$ is:
		\begin{align}\label{eq:19999911}
			\tilde{I}_{kk'} \defas \operatorname{var}\left(\sum_{m=1}^{M} a_{1m} {w}_{km} h_{k'm} \right),
		\end{align}
		and the distortion variance
		\begin{align}
			D \defas \operatorname{var}\left( \sum_{m=1}^{M} w_{km} d_m[n] \right).\label{eq:10857332}
		\end{align}
	\end{corollary}
	
	A system with perfectly linear \LNAs, where all first-degree coefficients are equal, $a_{1m} = 1$ for all $m$, and the distortion is zero, $D = 0$, is considered for comparison.  When this system uses maximum-ratio combining, i.e.\ $w_{km} = a^*_{1m} h^*_{km}$, and the channel is i.i.d.\ across $m$ then the gain and interference are given by:  
	\begin{align}
		G_k &\defas |g_k|^2 = \left|\sum_{m=1}^{M} \Exp\left[|h_{km}|^2\right]\right|^2 = M^2,\\
		\bar{I}_{kk'} &\defas \tilde{I}_{kk'} = M \operatorname{var}\left( h^*_{km} h_{k'm} \right).\label{eq:19988281822}
	\end{align}
	If it is assumed that the fading is Gaussian, i.e.\ such that $h_{km} \sim \CN(0,1)$, and the channel coefficients are independent across $m$ and $k$, then the interference variances are $\bar{I}_{kk'} = M$.  
	
	To get some qualitative insights, operation of the \LNAs at a fixed point, i.e.\ where $\{a_{1m}\}$ are fixed, is assumed in the following theorem.  Since the coefficients depend on the received power, such operation might be difficult in practice.  However, this mode of operation can be achieved approximately by varying the supply current to the amplifier to adjust for fading.  Deterministic coefficients can also appear in a fading environment, where the energy of the channel is constant, such as in highly frequency-selective channels or in free-space line-of-sight channels.  If the \LNA operation can be described by fixed gains $\{a_{1 m}\}$, the effective \SINR with nonlinear \LNAs is given by the following theorem.
	\begin{theorem}\label{the:891856178960781}
		In a system with \LNAs whose first-degree coefficients $\{a_{1m}\}$ are made constant and the channels $\{h_{km}\}$ are identically distributed for all $k$ and independent across $m$, the effective \SINR of a decoder using maximum-ratio combining $w_{km} = a^*_{1m} h^*_{km}$ is given by:
		\begin{align}
			\SINR_k = \frac{\rho P_k G_k}{\sum_{k'=1}^{K} P_{k'} \bar{I}_{kk'} + M N_0/T + D'},
		\end{align}
		where the gain loss is
		\begin{align}\label{eq:1907317}
			\rho \defas \frac{\left|\sum_{m=1}^{M} |a_{1m}|^2\right|^2}{M \sum_{m=1}^{M} |a_{1m}|^4}.
		\end{align}
		and the distortion power
		\begin{align}\label{eq:790019}
			D' \defas \frac{MD}{\sum_{m=1}^{M}|a_{1m}|^4}.
		\end{align}
	\end{theorem}
	The proof of this theorem is given in Appendix~\ref{app:29038}.  
	
	It is noted that the nonlinear \LNAs affect the rate $R_k$ of this linear decoder in two ways:
	\begin{itemize}
		\item[a)] There is a gain loss $\rho$ that is due to variations in the power amplifier.  Because of Cauchy-Schwartz inequality, $\rho \leq 1$, where equality occurs when all linear gains are the same, $a_{1m} = a_{1m'}$ for all $m$ and $m'$.  Hence, differences in linearity between different amplifiers lead to a somewhat reduced decoding gain.
		
		\item[b)] The distortion enters the rate expression in the same way as additional noise, and leads to an additional term in the denominator $D'$, which is proportional to the variance of the processed distortion $D$.
	\end{itemize}
	
	The phenomenon called \emph{desensitization} \cite[Ch.~2.1.1]{razavi1998rf} can be observed in the gains $\{a_{1m}\}$ of the desired linear part of the signal.  In practical amplifiers, these gains grow smaller the higher power the input signal has.  For example, from \eqref{eq:53836351982736}, it can be seen that a linearity of order $\Pi = 3$ has a gain that is given by:
	\begin{align}
		a_{1m} = b_{1m} + 2P^{\text{rx}}_m b_{3m},
	\end{align}
	where $P^{\text{rx}}_m$ is the received power at antenna $m$.  To model an amplifier with transfer characteristics that saturate, the complex coefficients $b_{1m}$ and $b_{3m}$ have opposite phases, which normally is the case.  Then the gain $|a_{1m}|^2$ can become small if the received power $P^{\text{rx}}_m$ is large.  
	
\begin{figure}
		\ifdraft\pgfplotsset{height=20ex,}\else\pgfplotsset{height=\flatFiguresHeight}\fi
		\centering
		\begin{tikzpicture}
		\begin{axis}[
		xmin=-10,
		xmax=0,
		ymax=0,
		xlabel={Received power relative to one-dB compression point $P^{\text{rx}}_m/P_{\text{1dB}}$ [dB]},
		ylabel={Gain $|a_{1m}|^2$ [dB]},
		grid=major,
		]
		\addplot[no marks,thick] coordinates {
			( -10.0 , -0.110268446603 )
			( -9.89898989899 , -0.114179722518 )
			( -9.79797979798 , -0.118225420889 )
			( -9.69696969697 , -0.122409446945 )
			( -9.59595959596 , -0.126735759176 )
			( -9.49494949495 , -0.131208365159 )
			( -9.39393939394 , -0.135831316952 )
			( -9.29292929293 , -0.140608706032 )
			( -9.19191919192 , -0.145544657757 )
			( -9.09090909091 , -0.150643325356 )
			( -8.9898989899 , -0.155908883433 )
			( -8.88888888889 , -0.161345520974 )
			( -8.78787878788 , -0.166957433879 )
			( -8.68686868687 , -0.172748817003 )
			( -8.58585858586 , -0.178723855743 )
			( -8.48484848485 , -0.18488671718 )
			( -8.38383838384 , -0.191241540818 )
			( -8.28282828283 , -0.197792428956 )
			( -8.18181818182 , -0.204543436766 )
			( -8.08080808081 , -0.211498562136 )
			( -7.9797979798 , -0.218661735381 )
			( -7.87878787879 , -0.226036808933 )
			( -7.77777777778 , -0.233627547152 )
			( -7.67676767677 , -0.24143761641 )
			( -7.57575757576 , -0.249470575671 )
			( -7.47474747475 , -0.257729867783 )
			( -7.37373737374 , -0.26621881176 )
			( -7.27272727273 , -0.274940596392 )
			( -7.17171717172 , -0.283898275554 )
			( -7.07070707071 , -0.29309476565 )
			( -6.9696969697 , -0.302532845723 )
			( -6.86868686869 , -0.312215160799 )
			( -6.76767676768 , -0.322144229168 )
			( -6.66666666667 , -0.332322454373 )
			( -6.56565656566 , -0.342752142812 )
			( -6.46464646465 , -0.35343552799 )
			( -6.36363636364 , -0.364374802578 )
			( -6.26262626263 , -0.375572159648 )
			( -6.16161616162 , -0.38702984459 )
			( -6.06060606061 , -0.398750219467 )
			( -5.9595959596 , -0.410735841766 )
			( -5.85858585859 , -0.422989559798 )
			( -5.75757575758 , -0.435514627282 )
			( -5.65656565657 , -0.448314839968 )
			( -5.55555555556 , -0.461394697575 )
			( -5.45454545455 , -0.474759594685 )
			( -5.35353535354 , -0.488416044758 )
			( -5.25252525253 , -0.502371941928 )
			( -5.15151515152 , -0.516636865884 )
			( -5.05050505051 , -0.531222435761 )
			( -4.94949494949 , -0.546142719803 )
			( -4.84848484848 , -0.561414708347 )
			( -4.74747474747 , -0.57705885874 )
			( -4.64646464646 , -0.593099721848 )
			( -4.54545454545 , -0.609566661141 )
			( -4.44444444444 , -0.626494676783 )
			( -4.34343434343 , -0.643925348811 )
			( -4.24242424242 , -0.661907915489 )
			( -4.14141414141 , -0.680500505115 )
			( -4.0404040404 , -0.699771542227 )
			( -3.93939393939 , -0.719801352268 )
			( -3.83838383838 , -0.740683992383 )
			( -3.73737373737 , -0.762529340434 )
			( -3.63636363636 , -0.785465479522 )
			( -3.53535353535 , -0.809641421603 )
			( -3.43434343434 , -0.835230221486 )
			( -3.33333333333 , -0.862432541862 )
			( -3.23232323232 , -0.891480741586 )
			( -3.13131313131 , -0.922643573823 )
			( -3.0303030303 , -0.956231598611 )
			( -2.92929292929 , -0.992603437016 )
			( -2.82828282828 , -1.03217302276 )
			( -2.72727272727 , -1.07541804385 )
			( -2.62626262626 , -1.12288981381 )
			( -2.52525252525 , -1.1752248735 )
			( -2.42424242424 , -1.2331587042 )
			( -2.32323232323 , -1.29754203817 )
			( -2.22222222222 , -1.3693603931 )
			( -2.12121212121 , -1.44975764489 )
			( -2.0202020202 , -1.54006470855 )
			( -1.91919191919 , -1.64183474657 )
			( -1.81818181818 , -1.75688680916 )
			( -1.71717171717 , -1.88736049166 )
			( -1.61616161616 , -2.03578516253 )
			( -1.51515151515 , -2.20516871241 )
			( -1.41414141414 , -2.39911281999 )
			( -1.31313131313 , -2.62196477122 )
			( -1.21212121212 , -2.87902046196 )
			( -1.11111111111 , -3.17680026007 )
			( -1.0101010101 , -3.52343035814 )
			( -0.909090909091 , -3.92917943421 )
			( -0.808080808081 , -4.40722733185 )
			( -0.707070707071 , -4.97478326963 )
			( -0.606060606061 , -5.65472611365 )
			( -0.505050505051 , -6.47798050821 )
			( -0.40404040404 , -7.48669643229 )
			( -0.30303030303 , -8.73707341192 )
			( -0.20202020202 , -10.2938136062 )
			( -0.10101010101 , -12.1725208025 )
			( 0.0 , -14.036398436 )
			};
		\end{axis}
		\end{tikzpicture}
		\caption{Desensitization of the linear gain $|a_{1m}|^2$ in the amplifier that is described by the polynomial coefficients in Table~\ref{tab:amp_coeffs}.}
		\label{fig:desensitization}
	\end{figure}
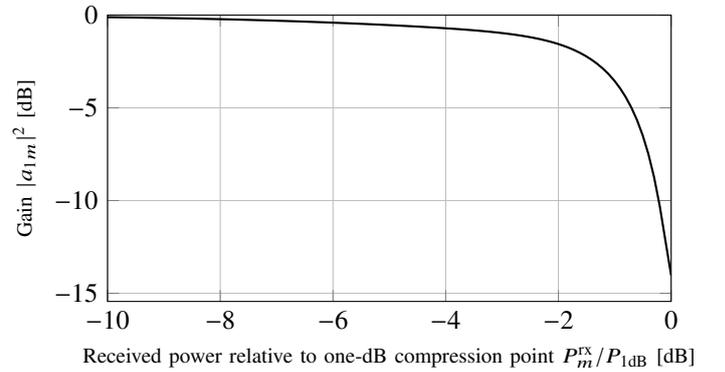

	In Figure~\ref{fig:desensitization}, the gain is shown for different amounts of received powers for a specific amplifier.  It can be observed that the desensitization effect can be significant.  Even if it turns out that the distortion $D'$ can be handled, desensitization will still have to be avoided if the \LNAs are to be operated close to saturation.
	
	Since the desired signal, as well as the interference and the thermal noise, all are amplified by the linear gains given by $\{|a_{1m}|^2\}$, their relative powers do not change significantly by the desensitization.  Desensitization however, increases the relative significance of the distortion, which is seen in \eqref{eq:790019}, where the denominator $\sum_{m=1}^{M} |a_{1m}|^4/M$ becomes small in case of desensitization.  
	
	Since the Hermite coefficients depend on the input power, which, in turn, depends on the channel fading, it is difficult to compute the use-and-then-forget bound in closed form in general.  As described in Theorem~\ref{the:891856178960781}, however, the overall effect of nonlinear \LNAs is a lower decoding gain and extra additive distortion.  The extra additive distortion will be studied in the following sections.

	\section{Spectral Analysis of Symbol Estimates}
	To gain insight into how large the error due to the nonlinear distortion in \eqref{eq:9777662553792} is, its second-order statistics will be analyzed using the correlation property of the Itô-Hermite polynomials.
	
	Two cyclostationary signals $x_1(t)$ and $x_2(t)$ with period $T$ have a cross-correlation function that is given by
	\begin{align}
		R_{x_1x_2}(t, \tau) \defas \Exp\bigl[x_1(t) x_2^*(t-\tau) \bigr]
	\end{align}
	that is periodic in the argument $t$ with period $T$.  By expanding the periodic cross-correlation function as a Fourier series, the cross-correlation of cycle index $\alpha$ can be obtained as:
	\begin{align}\label{eq:91001937}
		R^{(\alpha)}_{x_1x_2}(\tau) \defas \frac{1}{T} \int_{0}^{T} R_{x_1x_2}(t,\tau) e^{-j2\pi\alpha t/T}\symrm{d}t, \quad \alpha= 0, \pm 1, \pm 2, \ldots.
	\end{align}
	A good introduction to the second-order statistics of cyclostationary signals is given in \cite{gardner2006cyclostationarity}.  Similarly, two weak-sense stationary, discrete signals $y_1[n]$ and $y_2[n]$ have a cross-correlation function that is given by
	\begin{align}
		R_{y_1y_2}[\ell] \defas \Exp\bigl[y_1[n] y_2^*[n-\ell]\bigr].
	\end{align}
	
	The transmitted symbols are assumed to be independent and the cross-correlation function is therefore:
	\begin{align}
		R_{s_ks_{k'}}[\ell] = \delta[k-k'] \delta[\ell].
	\end{align}
	It then follows that the transmitted signals also are independent, $R_{x_kx_{k'}}(t, \tau) = 0$ for $k \neq k'$, and their cyclostationary autocorrelation functions are given by:
	\begin{align}
		R_{x_kx_{k}}(t, \tau) = \begin{cases}
			P_k \gamma(t, \tau) e^{-j2\pi B \tau}, &\text{if } k = K+1,\\
			P_k \gamma(t, \tau), &\text{otherwise},
		\end{cases}
	\end{align}
	where the aggregate pulse is
	\begin{align}
		\gamma(t, \tau) &\defas \sum_{n=-\infty}^{\infty} p(t-nT) p^*(t-nT-\tau).
	\end{align}
	The cross-correlation of the received signals is then:
	\begin{align}
		R_{u_mu_{m'}}(t, \tau) = \sum_{k=1}^{K+1} h_{km} h^*_{km'} R_{x_kx_{k}}(t, \tau).
	\end{align}
	Note that the channel is treated as deterministic, since we are analyzing the signal for a given channel realization.  Using the orthogonality property in \eqref{eq:910661188823} and the expansion of the amplifiers $\symcal{A}_m$ in \eqref{eq:00919922}, the cross-correlation of the output signals is
	\begin{align}
		R_{y_my_{m'}}(t,\tau) = a_{1m}a^*_{1m'} R_{u_mu_{m'}}(t,\tau) + R_{d_md_{m'}}(t,\tau),
	\end{align}
	where the cross-correlation of the distortion is:
	\begin{align}
		&R_{d_md_{m'}}(t,\tau) = \sum_{3\leq\varpi\leq\Pi:\text{odd}} a_{\varpi m}a^*_{\varpi m'} R_{u_{\varpi m}u_{\varpi m'}}(t,\tau),\label{eq:589}\\
		&R_{u_{\varpi m}u_{\varpi m'}}(t,\tau) \notag\\
		&\quad\defas \left(\frac{\varpi+1}{2}\right)! \left(\frac{\varpi-1}{2}\right)! R_{u_{m}u_{m'}}(t,\tau) |R_{u_{m}u_{m'}}(t,\tau)|^{\varpi-1}.\label{eq:109}
	\end{align}
	
	The amplified received signal is fed through a matched filter with impulse response $p^*(-\tau)/T$ prior to sampling.  The cross-correlation of cycle index $\alpha$ of the terms in \eqref{eq:91828988} is:
	\begin{align}
		R^{(\alpha)}_{\bar{u}_{\varpi m}\bar{u}_{\varpi m'}}(\tau) = \frac{1}{T^2} \left(R^{(\alpha)}_{u_{\varpi m}u_{\varpi m'}}(t) \star \gamma^{(\alpha)}(t) \right)(\tau),
	\end{align}
	where the cross-correlations of cycle index $\alpha$ of the input signal and the aggregate pulse are given by \eqref{eq:91001937} as:
	\begin{align}\label{eq:91003991}
		R^{(\alpha)}_{u_{\varpi m}u_{\varpi m'}}(\tau) &= \frac{1}{T} \int_{0}^{T} R_{u_{\varpi m}u_{\varpi m'}}(t, \tau) e^{-j2\pi\alpha t/T} \symrm{d}t,\\
		\gamma^{(\alpha)}(\tau) &\defas \frac{1}{T} \int_{0}^{T} \gamma(t,\tau) e^{-j2\pi t \alpha /T} \symrm{d}t\\
		&=\frac{1}{T} \left(p^*(-t) \star p(t) e^{-j2\pi t \alpha /T}\right)(\tau).
	\end{align}
	
	The filter output is then sampled to produce the discrete-time signal $u_{\varpi m}[n] \defas \bar{u}_{\varpi m}(t_0 + nT)$.  Since the sampling period $T$ is equal to the period of the cyclostationary continuous-time signal, the discrete-time signal is a weak-sense stationary signal with cross-correlation:
	\begin{align}
		R_{u_{\varpi m}u_{\varpi m'}}[\ell] &= R_{\bar{u}_{\varpi m}\bar{u}_{\varpi m'}}(t_0 + nT, \ell T)\\
		&= R_{\bar{u}_{\varpi m}\bar{u}_{\varpi m'}}(t_0, \ell T)\label{eq:0928556}\\
		&= \sum_{\alpha=-\infty}^{\infty} R^{(\alpha)}_{\bar{u}_{\varpi m}\bar{u}_{\varpi m'}}(\ell T) e^{j2\pi \alpha t_0/T}.\label{eq:781010}
	\end{align}
	In \eqref{eq:0928556}, the periodicity of the cross-correlation in its first argument is used.  In \eqref{eq:781010}, the periodic cross-correlation is expanded as a Fourier series.  It is noted that the sampling offset in \eqref{eq:0817176} was assumed to be $t_0 = 0$, which makes the complex exponentials in \eqref{eq:781010} equal to one for all $\alpha$.

	\section{Analysis of Third-Degree Distortion}
	In order to obtain some insights into the effect of the additive distortion on the decoding in an accessible way, the system is assumed to be noise-free, i.e.\ $z_m(t) = 0$, for all $m$, and only the third-degree term of the distortion
	\begin{align}
		d_{m}(t) = a_{3m} u_{3 m}(t)
	\end{align}
	will be studied.  This term is often the dominant one in the sense that it describes most of the distortion in and immediately around the frequency band of the desired signal.  The analysis of higher-degree terms can be done in a similar, albeit, more tedious way.  
	
	The cross-correlation of the third-degree distortion was given in \eqref{eq:589} and \eqref{eq:109} in terms of the third-degree cross-correlation of the received signal:
	\begin{align}
		&R_{u_{3m}u_{3m'}}(t,\tau) = 2 R_{u_{m}u_{m'}}(t,\tau)|R_{u_mu_{m'}}(t,\tau)|^2\\
		&\quad= 2\sum_{k=1}^{K+1} \sum_{k'=1}^{K+1} \sum_{k''=1}^{K+1} \bar{h}_{kk'k''m} \bar{h}^*_{kk'k''m'} P_k P_{k'} P_{k''} \gamma_{3,\nu(kk'k'')}(t,\tau),\label{eq:039288}
	\end{align}
	where the shorthand $\bar{h}_{kk'k''m} \defas h_{km} h_{k'm} h^*_{k''m}$ is used and the third-degree pulse $\gamma_{3,\nu}(t,\tau)$ is a frequency shifted version of the product $\gamma(t,\tau)|\gamma(t,\tau)|^2$:
	\begin{align}\label{eq:910292}
		\gamma_{3,\nu}(t,\tau) &= \gamma(t,\tau) |\gamma(t,\tau)|^2 e^{j2\pi \nu B \tau}.
	\end{align}
	The frequency shift  $\nu B$ is a multiple of the carrier frequency of the blocker $B$.  The multiplicity is determined by which of the indices $k,k',k''$ that equals $K+1$ (the index of the blocker) in the following way:
	\begin{align}
		\nu(k,k',k'') \defas I(k) + I(k') - I(k'') \in \{-1,0,1,2\},
	\end{align}
	where $I(k) = 1$ if $k=K+1$ and $I(k) = 0$ otherwise.  In Table~\ref{tab:number_of_terms}, the number of terms in \eqref{eq:039288} belonging to a given frequency index $\nu$ is shown together with the number of those terms that include different powers of the received power from the blocker.  It is noted that, if there is no blocker, $P_{K+1} = 0$, only the $K^3$ terms that belong to the pulse $\gamma_{3,0}(t,\tau)$ are left.  
	
	\begin{table}
		\ifdraft\linespread{1}\fi
		\centering
		\caption{Number of terms affected by blocker\strut}
		\label{tab:number_of_terms}
		\hspace{-0em}\begin{tabular}{rllll}
			\toprule
			$\nu=$&--1&0&1&2\\
			\midrule
			total \# terms&$K^2$&$2K+K^3$&$1+2K^2$&$K$\\
			\# terms with $P^3_{\!\!K+1}$ &&&1&\\
			\# terms with $P^2_{\!\!K+1}$ &&$2K$&&$K$\\
			\# terms with $P_{\!\!K+1}^{\phantom{1}}$ &$K^2$&&$2K^2$&\\
			\bottomrule
		\end{tabular}
	\end{table}
	
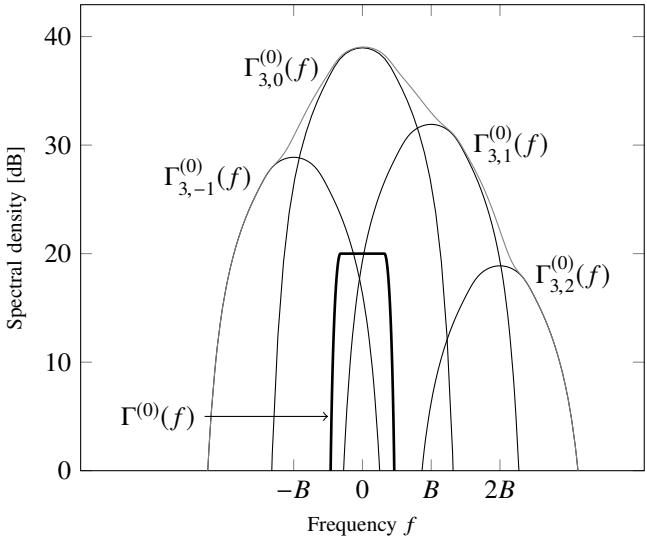
\begin{figure}
		\ifdraft\pgfplotsset{height=20ex,}\fi
		\centering
		\begin{tikzpicture}
		\begin{axis}[
		xmin=-5,
		xmax=5,
		ymin = 0,
		xlabel = {Frequency $f$},
		ylabel = {Spectral density [dB]},
		xtick={-1.22, 0, 1.22, 2.44},
		xticklabels={$-B$, 0, $B$, $2B$},
		restrict y to domain=-100:100,
		]
		
		\addplot[
		no marks,
		smooth,
		y filter/.code={\pgfmathparse{10*log10(#1*100)}\pgfmathresult},
		] table{
			-3.1	0
			-3	2.16E-08
			-2.9	5.65E-05
			-2.8	0.002572718
			-2.7	0.0248572532
			-2.6	0.105486643
			-2.5	0.2761444443
			-2.4	0.5460794301
			-2.3	0.9160094225
			-2.2	1.3859395813
			-2.1	1.9558697777
			-2	2.6258062457
			-1.9	3.3993745023
			-1.8	4.3089036282
			-1.7	5.3351254147
			-1.6	6.2606556783
			-1.5	6.9402653677
			-1.4	7.4009216515
			-1.3	7.6609210256
			-1.2	7.7209201126
			-1.1	7.5809194389
			-1	7.240912945
			-0.9	6.6972517324
			-0.8	5.9176647524
			-0.7	4.9214515637
			-0.6	3.9259218561
			-0.5	3.0764025166
			-0.4	2.3458269146
			-0.3	1.7158973319
			-0.2	1.1859678397
			-0.1	0.7560381872
			0	0.4261083165
			0.1	0.1961204755
			0.2	0.0636480141
			0.3	0.0113706197
			0.4	0.0007213888
			0.5	5.38E-06
			0.6	8.37E-13
			0.7	1.78E-15
			0.8	0
		};
		\node[anchor = east] at (axis cs: -1.8, 27) {$\Gamma^{(0)}_{3,-1}(f)$};
		
		\addplot[no marks,
		y filter/.code={\pgfmathparse{10*log10(#1*1020)}\pgfmathresult}
		] table{
			-1.9		0
			-1.8		8.04E-10
			-1.7		1.90E-05
			-1.6		0.0014089847
			-1.5		0.0171140536
			-1.4		0.0828061467
			-1.3		0.2341446686
			-1.2		0.4840938698
			-1.1		0.834023975
			-1		1.2839540344
			-0.9		1.8338840043
			-0.8		2.483814135
			-0.7		3.2354036447
			-0.6		4.1142268864
			-0.5		5.1292890369
			-0.4		6.0943435187
			-0.3		6.8232437254
			-0.2		7.3249200505
			-0.1		7.6249195662
			0		7.7249193433
			0.1		7.6249195662
			0.2		7.3249200505
			0.3		6.8232437254
			0.4		6.0943435187
			0.5		5.1292890369
			0.6		4.1142268864
			0.7		3.2354036447
			0.8		2.483814135
			0.9		1.8338840043
			1		1.2839540344
			1.1		0.834023975
			1.2		0.4840938698
			1.3		0.2341446686
			1.4		0.0828061467
			1.5		0.0171140536
			1.6		0.0014089847
			1.7		1.90E-05
			1.8		8.04E-10
			1.9		4.88E-15
			2		0
		};
		\node[anchor = east] at (axis cs: -.6, 37) {$\Gamma^{(0)}_{3,0}(f)$};
		
		\addplot[
		no marks,
		smooth,
		y filter/.code={\pgfmathparse{10*log10(#1*201)}\pgfmathresult},
		] table{
			-0.7			0
			-0.6			8.30E-13
			-0.5			5.38E-06
			-0.4			0.0007213888
			-0.3			0.0113706197
			-0.2			0.0636480141
			-0.1			0.1961204755
			0			0.4261083165
			0.1			0.7560381872
			0.2			1.1859678397
			0.3			1.7158973319
			0.4			2.3458269146
			0.5			3.0764025166
			0.6			3.9259218561
			0.7			4.9214515637
			0.8			5.9176647524
			0.9			6.6972517324
			1			7.240912945
			1.1			7.5809194389
			1.2			7.7209201126
			1.3			7.6609210256
			1.4			7.4009216515
			1.5			6.9402653677
			1.6			6.2606556783
			1.7			5.3351254147
			1.8			4.3089036282
			1.9			3.3993745023
			2			2.6258062457
			2.1			1.9558697777
			2.2			1.3859395813
			2.3			0.9160094225
			2.4			0.5460794301
			2.5			0.2761444443
			2.6			0.105486643
			2.7			0.0248572532
			2.8			0.002572718
			2.9			5.65E-05
			3			2.16E-08
			3.1			4.44E-15
			3.2			1.75E-15
			3.3			1.24E-15
			3.4			3.45E-15
			3.5			3.25E-15
			3.6			9.70E-16
			3.7			2.12E-15
			3.8			2.32E-15
			3.9			1.57E-15
			4			0
		};
		\node[anchor = west] at (axis cs: 1.8, 30) {$\Gamma^{(0)}_{3,1}(f)$};
		
		\addplot[
		no marks,
		smooth,
		y filter/.code={\pgfmathparse{10*log10(#1*10)}\pgfmathresult},
		] table{
			0.6				0
			0.7				1.23E-06
			0.8				0.0003412456
			0.9				0.0072599611
			1				0.047791982
			1.1				0.162042155
			1.2				0.3721221909
			1.3				0.6820518847
			1.4				1.091981316
			1.5				1.6019107187
			1.6				2.2118402344
			1.7				2.9219731334
			1.8				3.7441754219
			1.9				4.7140406427
			2				5.731309671
			2.1				6.5618084835
			2.2				7.1488719458
			2.3				7.5289208331
			2.4				7.7089215732
			2.5				7.6889220867
			2.6				7.4689220301
			2.7				7.0487134846
			2.8				6.4164143258
			2.9				5.54E+00
			3				4.51E+00
			3.1				3.57E+00
			3.2				2.77E+00
			3.3				2.08E+00
			3.4				1.49E+00
			3.5				1.00E+00
			3.6				6.12E-01
			3.7				3.22E-01
			3.8				1.32E-01
			3.9				3.50E-02
			4				0.0044326758
			4.1				0.0001469406
			4.2				2.09E-07
			4.3				3.44E-15
			4.4				0
		};
		\node[anchor = west] at (axis cs: 2.9, 18) {$\Gamma^{(0)}_{3,2}(f)$};
		
		\addplot[
		no marks,
		color=gray,
		smooth,
		y filter/.code={\pgfmathparse{10*log10(#1)}\pgfmathresult},
		] table {
			-5	4.7538671602648E-13
			-4.9	4.26471855896524E-13
			-4.8	2.0400348077535E-13
			-4.7	7.4676376193812E-13
			-4.6	8.7136894921808E-13
			-4.5	1.38844952584335E-12
			-4.4	3.3972824553576E-13
			-4.3	4.52970994047E-13
			-4.2	5.86445822459577E-13
			-4.1	9.05941988094E-13
			-4	1.67366120962269E-14
			-3.9	7.77156117238E-15
			-3.8	1.22124532709E-14
			-3.7	1.22124532709E-14
			-3.6	0
			-3.5	2.22044604925E-15
			-3.4	6.66133814775E-15
			-3.3	0
			-3.2	1.5570877920372E-13
			-3.1	2.22044604925E-15
			-3	2.16364177907045E-06
			-2.9	0.0056528967
			-2.8	0.257271801
			-2.7	2.4857253243
			-2.6	10.5486643002
			-2.5	27.6144444267
			-2.4	54.6079430124
			-2.3	91.6009422548
			-2.2	138.593958132
			-2.1	195.586977769
			-2	262.580624569
			-1.9	339.93745023
			-1.8	430.8903636384
			-1.7	533.5318943898
			-1.6	627.5027322531
			-1.5	711.4828714745
			-1.4	824.5544347433
			-1.3	1004.9196645575
			-1.2	1265.867758403
			-1.1	1608.7963983574
			-1	2033.7244096238
			-0.9	2540.2868576454
			-0.8	3125.2568929482
			-0.7	3792.2568739304
			-0.6	4589.1036096924
			-0.5	5539.5161513326
			-0.4	6450.9580797016
			-0.3	7133.5838275967
			-0.2	7602.8084862851
			-0.1	7892.4419918271
			0	8007.6763334319
			0.1	7948.9936807062
			0.2	7716.1627886693
			0.3	7305.7410255294
			0.4	6687.8137378037
			0.5	5850.2322617722
			0.6	4985.6217171533
			0.7	4289.3234941794
			0.8	3722.9444453944
			0.9	3216.7818822275
			1	2765.534536884
			1.1	2376.0896832343
			1.2	2049.4019116822
			1.3	1785.493206988
			1.4	1582.9673347049
			1.5	1428.4687808016
			1.6	1281.9473581028
			1.7	1101.5992926116
			1.8	903.5313843036
			1.9	730.4146813897
			2	585.1001520936
			2.1	458.7479101508
			2.2	350.0625753031
			2.3	259.4071022629
			2.4	186.8511811867
			2.5	132.3942541643
			2.6	95.892035544
			2.7	75.4834427483
			2.8	64.6812595783
			2.9	55.3766957443
			3	45.0926110602
			3.1	35.6879791675
			3.2	27.7183156674
			3.3	20.8185475811
			3.4	14.919248854
			3.5	10.0199508894
			3.6	6.1206545884
			3.7	3.2213487392
			3.8	1.3185818034
			3.9	0.3496694703
			4	0.044326758
			4.1	0.0014694056
			4.2	2.0870945259216E-06
			4.3	2.18186579914428E-13
			4.4	7.77156117238E-14
			4.5	1.11022302463E-14
			4.6	2.49980591782214E-13
			4.7	1.532107773985E-13
			4.8	1.354472090045E-13
			4.9	6.0895732900704E-13
		};
		
		\addplot[
		no marks,
		line width = 1pt,
		y filter/.code={\pgfmathparse{10*log10(#1)}\pgfmathresult},
		] table{
			-0.63 4.70191687469e-15
			-0.62 1.79186668374e-14
			-0.61 1.33462823413e-17
			-0.6 0.00259008901352
			-0.59 0.0410205044416
			-0.58 0.2041594419
			-0.57 0.630011065127
			-0.56 1.4914568117
			-0.55 2.97802787975
			-0.54 5.27527152177
			-0.53 8.54349023108
			-0.52 12.8977030477
			-0.51 18.3905959566
			-0.5 25.0000000256
			-0.49 32.6220797482
			-0.48 41.0709586579
			-0.47 50.084991414
			-0.46 59.3393530979
			-0.45 68.4641010381
			-0.44 77.0664139199
			-0.43 84.7553638867
			-0.42 91.1673582944
			-0.41 95.9903167439
			-0.4 98.9847318496
			-0.39 100.000000177
			-0.38 100.000002781
			-0.37 100.000001476
			-0.36 100.000001037
			-0.35 100.000000816
			-0.3 100.00000044
			-0.2 100.000000276
			-0.1 100.00000023
			0.0 100.000000218
			0.2 100.000000276
			0.3 100.00000044
			0.35 100.000000816
			0.36 100.000001037
			0.37 100.000001476
			0.38 100.000002781
			0.39 100.000000177
			0.4 98.9847318496
			0.41 95.9903167439
			0.42 91.1673582944
			0.43 84.7553638867
			0.44 77.0664139199
			0.45 68.4641010381
			0.46 59.3393530979
			0.47 50.084991414
			0.48 41.0709586579
			0.49 32.6220797482
			0.5 25.0000000256
			0.51 18.3905959566
			0.52 12.8977030477
			0.53 8.54349023108
			0.54 5.27527152177
			0.55 2.97802787975
			0.56 1.4914568117
			0.57 0.630011065127
			0.58 0.2041594419
			0.59 0.0410205044417
			0.6 0.00259008901352
			0.61 1.33460446275e-17
			0.62 1.79186605173e-14
			0.63 4.70191242733e-15
			0.64 2.17172487609e-15
		};
		\node[anchor=east] at (axis cs: -2.8, 5) {$\Gamma^{(0)}(f)$};
		\draw (axis cs: -2.8, 5) edge[->] (axis cs: -0.61, 5);
		\end{axis}
		\end{tikzpicture}
		\caption{The four pulse shapes in the sum in \eqref{eq:039288}.  The pulse $p(\tau)$ has been chosen as a root-raised cosine with roll-off 0.22, and the Fourier transform $\Gamma^{(0)}(f)$ of its ambiguity function at cycle index zero is shown for comparison.  The pulses have been scaled by a factor equal to the number of terms corresponding to each pulse when there are $K = 10$ served users, see Table~\ref{tab:number_of_terms}; this however is an arbitrary scaling.  The sum of the pulses is shown in grey.}
		\label{fig:third_degree_pulse_types}
	\end{figure}

	The cross-correlation of cycle index $\alpha$ for the periodic correlation function in \eqref{eq:039288} is thus
	\begin{multline}
		R^{(\alpha)}_{u_{3m}u_{3m'}}(\tau) \\
		= 2\sum_{k=1}^{K+1} \sum_{k'=1}^{K+1} \sum_{k''=1}^{K+1} \bar{h}_{kk'k''m} \bar{h}^*_{kk'k''m'} P_k P_{k'} P_{k''} \gamma_{3,\nu(kk'k'')}(\tau),
	\end{multline}
	where the pulses are given by:
	\begin{align}
		\gamma_{3,\nu}^{(\alpha)}(\tau) \defas \frac{1}{T} \int_{0}^{T} \gamma_{3,\nu}(t,\tau) e^{-j2\pi\alpha t/T} \symrm{d}t,\quad\nu=-1,0,1,2.
	\end{align}
	The Fourier transforms $\Gamma_{3,\nu}^{(0)}(f)$ of these pulses for $\alpha = 0$ are shown in Figure~\ref{fig:third_degree_pulse_types}.  
	
\begin{figure}
		\ifdraft\fi
		\newlength{\figbredd}
		\ifdraft
		\setlength{\figbredd}{.5\linewidth}
		\pgfplotsset{height=20ex,width=\figbredd}
		\else
		\setlength{\figbredd}{\linewidth}
		\pgfplotsset{height=\flatFiguresHeight}
		\fi
		\centering
		\begin{minipage}{\figbredd}
			\begin{tikzpicture}
			\begin{axis}[
			xlabel={Lag $\ell$ [samples]},
			ylabel={Correlation magnitude},
			ymin = 0,
			xmin=-10,
			xmax=10,
			legend pos=north east,
			legend cell align=left,
			legend style={font=\footnotesize},
			cycle list name=mark list,
			]
			\addplot+[ycomb] table {
				-10 0.00101849889315
				-9 0.00168685818627
				-8 0.00173861711037
				-7 0.000772666507798
				-6 0.00140105756984
				-5 0.00459948810109
				-4 0.0081167616662
				-3 0.010435244103
				-2 0.00749883490519
				-1 0.0277564387626
				0 0.713021134764
				1 0.027758230886
				2 0.00749791715803
				3 0.0104347651816
				4 0.00811658783408
				5 0.00459947639678
				6 0.00140109917246
				7 0.000772616950188
				8 0.00173857730929
				9 0.00168699270661
				10 0.00101853028576
			};
			\node[anchor=north west] at (axis cs: 0, 0.713021134764) {$|\gamma_{3,0}[\ell]|$};
			
			\addplot+[ycomb] table {
				-10 0.000225138675805
				-9 0.000376109450161
				-8 0.00040545886642
				-7 0.000240403664707
				-6 0.000842810762773
				-5 0.00253187987775
				-4 0.00555723157213
				-3 0.010733921946
				-2 0.0203702109363
				-1 0.0496184312493
				0 0.0826537216682
				1 0.0496186004928
				2 0.0203700310385
				3 0.0107336797895
				4 0.00555703928818
				5 0.002531762451
				6 0.000842755671715
				7 0.000240375249879
				8 0.000405430371674
				9 0.000376045470858
				10 0.000225104983649
			};
			\node[anchor=south west] at (axis cs: 0, 0.0826537216682) {$|\gamma_{3,1}[\ell]|$};
			\draw (axis cs: -.7, 0.0826537216682) edge[|-|] node[anchor=east] {$\approx$9.4\,dB} (axis cs: -.7, 0.713021134764);
			\end{axis}
			\end{tikzpicture}\\[\abovecaptionskip]\centering\footnotesize
			a) Ambiguity functions\ifdraft\else\bigskip\fi
		\end{minipage}\ifdraft\else\pgfplotsset{height=\flatFiguresHeight}\newline\fi%
		\begin{minipage}{\figbredd}
			\begin{tikzpicture}
			\begin{axis}[
			xlabel={Normalized frequency $\theta$},
			ylabel={Spectral density [dB]},
			xmin=-.5,
			xmax=.5,
			y filter/.code={\pgfmathparse{10*log10(\pgfmathresult))}},
			]
			\addplot[
			no marks,
			smooth,
			] table {
				-0.5 0.7725037712863232
				-0.49 0.7724037901076448
				-0.48 0.7721038404327145
				-0.47 0.7716039055726761
				-0.46 0.7709039630272276
				-0.45 0.7700039909912576
				-0.44 0.7689039748679547
				-0.43 0.7676039119239456
				-0.42 0.7661038126053034
				-0.41 0.764403697867489
				-0.4 0.7625035929989112
				-0.39 0.7604035195637583
				-0.38 0.7581034879149543
				-0.37 0.7556034929088011
				-0.36 0.7529035147825345
				-0.35 0.7500035256330564
				-0.34 0.7469034998041574
				-0.33 0.7436034242133951
				-0.32 0.7401033028091873
				-0.31 0.7364031484949504
				-0.3 0.7325029563809874
				-0.29 0.7284026542493061
				-0.28 0.7241020293533315
				-0.27 0.7196006344971013
				-0.26 0.7148976799869668
				-0.25 0.7099919208262344
				-0.24 0.704881549990772
				-0.23 0.6995641087170582
				-0.22 0.6940364237622739
				-0.21 0.688294580125879
				-0.2 0.6823339363987825
				-0.19 0.6761491893000906
				-0.18 0.6697344946043919
				-0.17 0.6630836546719125
				-0.16 0.6561903926110907
				-0.15 0.6490487702009708
				-0.14 0.6416539824548344
				-0.13 0.63400490785398
				-0.12 0.6261231394671867
				-0.11 0.6192337649738597
				-0.1 0.6354750513195557
				-0.09 0.651077158191907
				-0.08 0.6648241232753528
				-0.07 0.6767322471628868
				-0.06 0.6868558113752818
				-0.05 0.6952614070115424
				-0.04 0.702018925516477
				-0.03 0.7071948860723659
				-0.02 0.7108469426912599
				-0.01 0.7130195002457942
				0.0 0.7137405075512465
				0.01 0.713019500244713
				0.02 0.7108469426890967
				0.03 0.7071948860691253
				0.04 0.7020189255121819
				0.05 0.6952614070062518
				0.06 0.6868558113691078
				0.07 0.6767322471560115
				0.08 0.6648241232680393
				0.09 0.6510771581845006
				0.1 0.6354750513124958
				0.11 0.6192337649678894
				0.12 0.6261231394624559
				0.13 0.634004907850026
				0.14 0.6416539824514556
				0.15 0.649048770198001
				0.16 0.6561903926083912
				0.17 0.6630836546693865
				0.18 0.6697344946019969
				0.19 0.6761491892978355
				0.2 0.6823339363967124
				0.21 0.6882945801240498
				0.22 0.6940364237607266
				0.23 0.6995641087157971
				0.24 0.7048815499897564
				0.25 0.7099919208253852
				0.26 0.7148976799861838
				0.27 0.7196006344962891
				0.28 0.7241020293524255
				0.29 0.7284026542482896
				0.3 0.7325029563798965
				0.31 0.7364031484938613
				0.32 0.7401033028081944
				0.33 0.7436034242125771
				0.34 0.7469034998035511
				0.35 0.750003525632642
				0.36 0.7529035147822439
				0.37 0.7556034929085406
				0.38 0.7581034879146388
				0.39 0.7604035195633377
				0.4 0.7625035929983849
				0.41 0.7644036978669038
				0.42 0.7661038126047368
				0.43 0.7676039119234792
				0.44 0.768903974867649
				0.45 0.770003990991134
				0.46 0.7709039630272618
				0.47 0.7716039055728066
				0.48 0.7721038404328631
				0.49 0.7724037901077401
				0.5 0.7725037712863232
			};
			\node at (axis cs: -.3, -9) {$\Gamma_{3,0}[\theta]$};
			
			\addplot[
			no marks,
			smooth,
			] table{
				-0.5 0.15382921909943642
				-0.49 0.1353280198506976
				-0.48 0.1162182672010808
				-0.47 0.09698293363527617
				-0.46 0.07813144869017086
				-0.45 0.060189056645015906
				-0.44 0.04368543562010899
				-0.43 0.029142825321236687
				-0.42 0.017063920874897585
				-0.41 0.00791980521125433
				-0.4 0.0021382864465234463
				-0.39 8.72228697435945e-05
				-0.38 0.00012289401685289247
				-0.37 0.00017776439922972953
				-0.36 0.0002452154772453058
				-0.35 0.0003292455476591599
				-0.34 0.00043345787712742445
				-0.33 0.0005615566735671096
				-0.32 0.0007174547299515638
				-0.31 0.0009052834203768373
				-0.3 0.0011293687522428153
				-0.29 0.0013941914466299304
				-0.28 0.0017043380881476278
				-0.27 0.002064447216435325
				-0.26 0.0024791531246407816
				-0.25 0.0029530296275072955
				-0.24 0.0034905357405400662
				-0.23 0.004095964947479269
				-0.22 0.004773399495071499
				-0.21 0.005526670936764797
				-0.2 0.006359327944174986
				-0.19 0.007274612199376848
				-0.18 0.008275442943368454
				-0.17 0.009364410454725969
				-0.16 0.010543778346718574
				-0.15 0.011815494106381982
				-0.14 0.013181206802846818
				-0.13 0.014642290452288848
				-0.12 0.016199871227772875
				-0.11 0.01785485661052834
				-0.1 0.019607964727972896
				-0.09 0.02145975248935153
				-0.08 0.023410641631151127
				-0.07 0.0254609423024149
				-0.06 0.027610874231613472
				-0.05 0.029860585733055306
				-0.04 0.032210170806762875
				-0.03 0.03465968440877311
				-0.02 0.03720915572329746
				-0.01 0.039858599079321155
				0.0 0.04260802212377995
				0.01 0.04545743103296053
				0.02 0.04840683287863877
				0.03 0.0514562356660814
				0.04 0.05460564689728608
				0.05 0.05785507166950507
				0.06 0.0612045112363447
				0.07 0.06465396265287165
				0.08 0.06820341968527473
				0.09 0.07185287472106788
				0.1 0.07560232109373699
				0.11 0.0794517551171452
				0.12 0.08340117722814504
				0.13 0.08745059191378475
				0.14 0.09160000645891792
				0.15 0.09584942888319395
				0.16 0.10019886565358718
				0.17 0.10464831981025817
				0.18 0.10919779002736806
				0.19 0.11384727088449426
				0.2 0.11859675431172667
				0.21 0.12344623186425512
				0.22 0.12839569724797123
				0.23 0.13344514841089858
				0.24 0.1385945885710901
				0.25 0.14384402577760508
				0.26 0.14919347096925023
				0.27 0.15464293493614473
				0.28 0.1601924249937573
				0.29 0.1658419424201113
				0.3 0.17159148167266175
				0.31 0.17744103204303266
				0.32 0.1833905817859573
				0.33 0.18944012408179606
				0.34 0.19558966386875204
				0.35 0.20183922544171
				0.36 0.20818886522019525
				0.37 0.21463871360943304
				0.38 0.2211891711374538
				0.39 0.22783628225668956
				0.4 0.2333893759800946
				0.41 0.23654379363770667
				0.42 0.23716059112320714
				0.43 0.23515875903256322
				0.44 0.23051528476683308
				0.45 0.22326891487657297
				0.46 0.21352242009308578
				0.47 0.20144335103233085
				0.48 0.1872632141360695
				0.49 0.17127502242831044
				0.5 0.15382921909943642
			};
			\node[anchor=west] at (axis cs: -.1, -22) {$\Gamma_{3,1}[\theta]$};
			\end{axis}
			\end{tikzpicture}\\[\abovecaptionskip]\centering\footnotesize
			b) Fourier transform
		\end{minipage}
			\ifdraft\vspace{1ex}\fi
			\caption{Time and frequency characterizations of the ambiguity functions $\gamma_{3,\nu}[\ell]$.  The pulse $p(\tau)$ has been chosen as a root-raised cosine pulse with roll-off 0.22.  Figure~a: The magnitude of $|\gamma_{3,\nu}[\ell]|$ for $\nu = 0, 1$.  Figure~b: The discrete-time Fourier transforms $\Gamma_{3,\nu}[\theta]$ of $\gamma_{3,\nu}[\ell]$. Note that $|\gamma_{3,-1}[\ell]| = |\gamma_{3,1}[\ell]|$ and $\Gamma_{3,-1}[\theta] = \Gamma_{3,1}[-\theta]$.}
		\label{fig:correlation_terms}
	\end{figure}
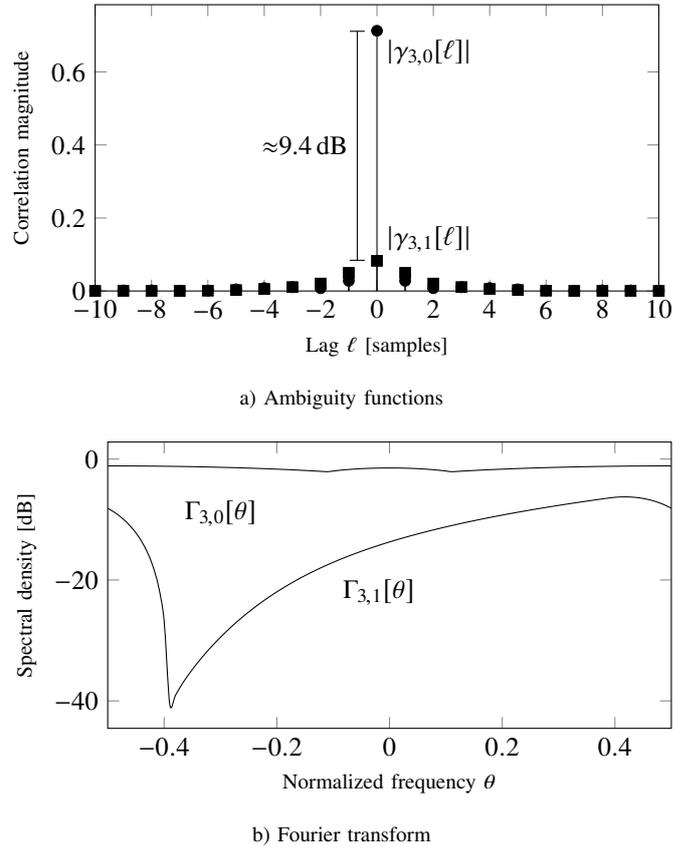

	The cross-correlation of the matched-filtered and sampled signal that was given in \eqref{eq:781010} can now be written as:
	\begin{align}
		&R_{u_{3m}u_{3m'}}[\ell] = a_{3m} a^*_{3m'} \!\! \sum_{\alpha=-\infty}^{\infty} {\left(R^{(\alpha)}_{u_{3m}u_{3m'}}(\tau) \star \gamma^{(\alpha)}(\tau) \right)}(t)\Bigr|_{t=\ell T}\\
		&= a_{3m} a^*_{3m'} \sum_{k=1}^{K+1} \sum_{k'=1}^{K+1} \sum_{k''=1}^{K+1} \bar{h}_{kk'k''m} \bar{h}^*_{kk'k''m'} P_k P_{k'} P_{k''} \gamma_{3,\nu(k,k',k'')}[\ell]\label{eq:03672}\\
		&=a_{3m} a^*_{3m'} \sum_{\mathclap{\nu=-1}}^{1} \gamma_{3,\nu}[\ell] \sum_{\mathclap{(k,k',k'') \in \symcal{K}_\nu}} P_k P_{k'} P_{k''} \bar{h}_{kk'k''m} \bar{h}^*_{kk'k''m'},\label{eq:90019}
	\end{align}
	where the sets $\symcal{K}_\nu$ contain the user indices that affect a given pulse:
	\begin{align}
		\symcal{K}_{\nu} &\defas \{(k,k',k''): \nu(k,k',k'') = \nu\}.
	\end{align}
	and the three ambiguity functions are defined as follows:
	\begin{align}
		\gamma_{3,\nu}[\ell] \defas \frac{1}{T^2} \sum_{\alpha=-\infty}^{\infty} {\left(\gamma_{3,\nu}^{(\alpha)}(\tau) \star \gamma^{(\alpha)}(\tau) \right)}(t)\Bigr|_{t=\ell T}.
	\end{align}
	Since the pulses $\gamma^{(\alpha)}(\tau)$, whose spectrum is limited to $[-B/2,B/2]$ by construction, and $\gamma_{3,2}(\tau)$ have disjoint supports in the frequency domain, $\gamma_{3,2}[\ell] = 0$ for all $\ell$, as is seen in Figure~\ref{fig:third_degree_pulse_types}, where three of the pulses overlap with the receive filter.  The cross-correlation of the matched-filtered signal therefore only contains three pulses.  The corresponding three nonzero ambiguity functions can be seen in Figure~\ref{fig:correlation_terms} for a root-raised cosine pulse $p(\tau)$ with roll-off 0.22.  It can be seen that the center pulse $\gamma_{3,0}[\ell]$ is practically frequency flat, while the adjacent pulses $\gamma_{3,-1}[\ell]$ and $\gamma_{3,1}[\ell]$ are frequency selective.  The frequency content of these pulses mostly lies towards the low and high frequencies respectively, which means that distortion from these pulses can be avoided at certain frequencies.
	
	From \eqref{eq:90019}, it can be seen that the third-degree distortion term has a spatial pattern that is decided by the composite channels $\{\bar{h}_{kk'k''m}, m=1,\dots,M\}$.  If decoding is done as in \eqref{eq:8109293}, the distortion term will combine coherently for certain choices of $\{w_{km}, m=1,\ldots,M\}$ and destructively for others.  This is described by the autocorrelation function of the additive distortion term $e_k[n]$:
	\begin{align}
		R_{e_ke_k}[\ell] &= \sum_{m=1}^{M} \sum_{m'=1}^{M} w_{km} w^*_{km'} R_{u_{3m}u_{3m'}}[\ell]\\
		&= \sum_{\mathclap{\nu=-1}}^{1}\! \gamma_{3,\nu}[\ell] \sum_{\mathclap{(k,k',k'') \in \symcal{K}_\nu}} P_{k'}P_{k''}P_{k'''} \notag\\
		&\quad\times\sum_{m=1}^{M} \sum_{m'=1}^{M} \!\! a_{3m} a^*_{3m'} \bar{h}_{k'k''k'''m}\bar{h}^*_{k'k''k'''m'} w_{km} w^*_{km'}\label{eq:0091929}
	\end{align}
	
	The following observations can be made from \eqref{eq:0091929}.
	\begin{observation}
		It is seen that the distortion combines constructively in the directions given by 
		\begin{align}\label{eq:9272322}
		\{a_{3m}\bar{h}_{kk'k''m}, m=1,\dots,M\}
		\end{align}
		for $k,k',k''=1,\ldots,K$, which means that a user $\chi$ with a decoding vector $(w_{\chi 1}, w_{\chi 2}, \ldots, w_{\chi M})^\tr$ that is not orthogonal to all the distortion directions, e.g.\ the third-degree distortion vectors $(a_{31} \bar{h}^*_{kk'k''1}, a_{32} \bar{h}^*_{kk'k''2}, \ldots, a_{3M} \bar{h}^*_{kk'k''M})^\tr$ given by \eqref{eq:9272322}, will suffer from distortion.  
		The larger the inner product between decoding vector and one of the distortion directions, the larger the distortion will be.
	\end{observation}
	
	\begin{observation}
		When the directions in \eqref{eq:9272322} are not parallel to the channel of the user, the distortion can be mitigated by choosing decoding weights that make the sum in \eqref{eq:0091929} small.  Distortion mitigation would require the distortion directions to be established.  It remains to be shown, if it is possible to estimate the distortion directions, and the coefficients $\{a_{3m}\}$, sufficiently well in practice to perform such distortion mitigation.  
	\end{observation}
	
	\begin{observation}
		The number of directions $\{a_{3m}\bar{h}_{kk'k''m}, m=1,\dots,M\}$ is proportional to $K^3$.  The proportionality constant is smaller than one, since some of the distortion directions are the same, e.g.\ $\bar{h}_{kk'k''m} = \bar{h}_{k'kk''m}$.  When the number of directions is greater than the dimension of the signal space, which is $M$, and all directions have the same power $P_k P_{k'} P_{k''}$, the distortion can be isotropic and the distortion is picked up by any choice of decoding weights $\{w_{km}\}$.
	\end{observation}
	
	\section{Free-Space Line-of-Sight and Maximum-Ratio Combining}
	To develop an intuition for how the distortion affects the decoding, the special case of free-space line-of-sight channels and maximum-ratio combining is considered.  If user $k$ is located at a distance $d_{km}$ from antenna $m$ and the distances $\{d_{km}, m=1,\ldots,M\}$ are similar, so that the path loss to each antenna is the same, and the individual antenna gains are identical, then the frequency-flat channel to user $k$ from antenna $m$ is given by:
	\begin{align}
		h_{km} = e^{jd_{km}/\lambda},
	\end{align}
	where $\lambda$ is the wavelength of the signal.  Using this notation, the composite channel becomes:
	\begin{align}
		\bar{h}_{kk'k''m} = e^{j(d_{km} + d_{k'm} - d_{k''m})}.\label{eq:91077}
	\end{align}
	
	The nominal nonlinearity characteristics of the amplifier is an outcome of the amplifier design, rather than something that can be arbitrarily chosen.  Also, the typical goal of a manufacturer is to minimize variations between product samples.  We will therefore
	assume, in this section, that all the amplifiers are identical.  Since the modulus of all channel coefficients is the same, the received energy at all antennas is the same, and the Hermite coefficients $a_{3 m} = a_3$ are the same for all $m$.  
	
	It is also assumed that maximum-ratio combining is used, i.e.\ that
	\begin{align}\label{eq:907761788992}
	w_{km} = a^*_{1m} h_{km}^*, \quad \text{for all } k,m.
	\end{align}
	Just like the third-degree coefficients are independent of the antenna index $m$, the first-degree coefficients $\{a_{1m}\}$ are too.  For notational simplicity and without loss of generality, it will be assumed that $a_{1m} = 1$ for all $m$.
	
	Under these assumptions, we see that, among the spatial directions of the distortion in \eqref{eq:9272322}, there are directions that are identical to the spatial direction of the desired in-band signal.  
	For example, from \eqref{eq:91077} it is seen that $\bar{h}_{kk'k''m} = h_{km}$ when $k'=k''$ and thus that the direction of the distortion and of user $k$ are the same.  
	These distortion terms thus combine in the same way as the desired in-band signal.  
	The following more general conclusion can be made.
	
	\begin{observation}
		In free-space line-of-sight and with identical amplifiers, some of the distortion directions in \eqref{eq:9272322} are parallel to the channels of the users and combine coherently for any choice of decoding weights, for which the signal of interest combines coherently.  What is more, with a single user, the distortion only has one term, which has a direction identical to that of the desired signal, making distortion mitigation through linear processing impossible.  
	\end{observation}
	
	To gain further intuition, we study the   case of a uniform linear array with the users in its geometric far-field. The channel coefficients of user $k$ are   functions of the incident angle $\theta_k$ relative to the broadside of the array:
	\begin{align}
		h_{km} = e^{jm\phi_k},
	\end{align}
	where $\phi_k \defas -2\pi \sin(\theta_k) \Delta/\lambda$ is the normalized sine angle of the incident signal, $\lambda = c/f_c$ is the wavelength of the signal at the carrier frequency $f_c$ and $\Delta$ is the antenna spacing.  Using this notation, the composite channel becomes:
	\begin{align}
		\bar{h}_{kk'k''m} = e^{jm(\phi_k + \phi_{k'} - \phi_{k''})}.
	\end{align}
		
\begin{figure}
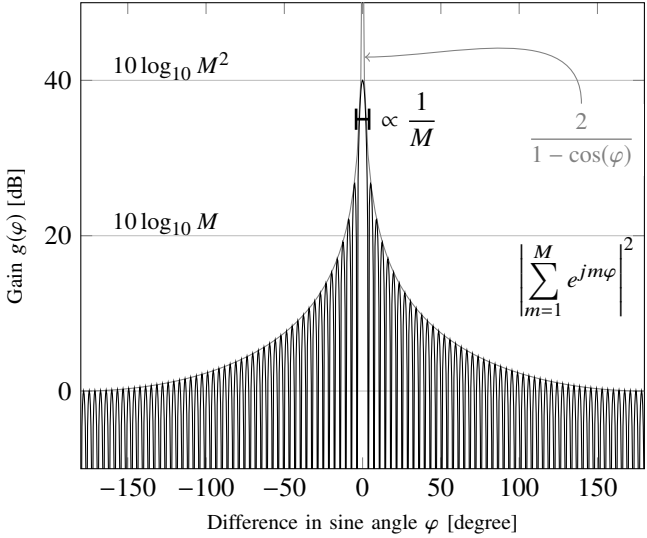

		\centering

		\caption{The double sum in \eqref{eq:8181818001} is approximately zero, except when $\varphi$ is close to zero.  At $\varphi=0$, the sum is equal to $M^2$.  Here the sum is evaluated for $M=100$.}
		\label{fig:approximate_sum}
	\end{figure}

	In this case, the inner sum of the distortion error in \eqref{eq:0091929} becomes
	\begin{align}
		&\sum_{m=1}^{M} \sum_{m'=1}^{M} \!\! a_{3m} a^*_{3m'} \bar{h}_{k'k''k'''m}\bar{h}^*_{k'k''k'''m'} w_{km} w^*_{km'}\notag\\
		&\quad= |a_{3}|^2 \sum_{m=1}^{M} \sum_{m'=1}^{M} \bar{h}_{k'k''k'''m}\bar{h}^*_{k'k''k'''m'} h^*_{km} h_{km'}\label{eq:155552939}\\
		&\quad= |a_{3}|^2 \sum_{m=1}^{M} \sum_{m'=1}^{M} e^{j(m - m')(\phi_{k'} + \phi_{k''} - \phi_{k'''} - \phi_{k})}.\label{eq:8181818001}\\
		&\quad= |a_3|^2 g(\phi_{k'} + \phi_{k''} - \phi_{k'''} -\phi_k),\label{eq:91097811}
	\end{align}
	where the double sum has been denoted by:
	\begin{align}
		g(\varphi) \defas \sum_{m=1}^{M} \sum_{m'=1}^{M} e^{j(m - m')\varphi} = \left|\sum_{m=1}^{M} e^{jm\varphi}\right|^2.\label{eq:1801020}
	\end{align}
	The factor $g(\phi_{k'} + \phi_{k''} - \phi_{k'''} -\phi_k)$ will be referred to as the \emph{array gain} of the $(k',k'',k''')$\mbox{-}th term in \eqref{eq:0091929}\footnote{Some times the array gain is scaled by $1/M$, so that the maximum array gain equals $M$, the number of antennas, and not $M^2$ as is the case here without the scaling.}.  It is shown for different phase differences $\varphi$ in Figure~\ref{fig:approximate_sum} for $M=100$.  From its definition in \eqref{eq:1801020}, the following well-known properties of the array gain can be observed:
	\begin{itemize}
		\item It has a main lobe around $\varphi = 0$ of width $2\pi/M$.
		
		\item It has a maximum at $\varphi = 0$, where $g(0) = M^2$.
		
		\item Its envelope is upper bounded by:
		\begin{align}\label{eq:83571910093}
			g(\varphi) \leq \psi(\varphi) \defas \frac{2}{1-\cos(\varphi)}.
		\end{align}
		
		\item Because of \eqref{eq:83571910093}, $g(\varphi)$ stays finite when $M$ grows for all $\varphi$ except $\varphi = 0$, for which $g(0) = M^2$.
	\end{itemize}
	For a small number of antennas, the width of the main lobe, $2\pi/M$, can be significant, e.g.\ with $M=16$ antennas, the width is \SI{22.5}{\degree} in sine angle.  For larger number of antennas, however, the main lobe is narrow, e.g.\ with $M=100$ antennas, the case shown in Figure~\ref{fig:approximate_sum}, the width is \SI{3.6}{\degree} in sine angle.

	The autocorrelation of the error due to the distortion in free-space line-of-sight and maximum-ratio combining is thus:
	\begin{align}
	R_{e_ke_k}[\ell] &= |a_3|^2 \sum_{\mathclap{\nu=-1}}^{1}\! \gamma_{3,\nu}[\ell] \sum_{\mathclap{(k,k',k'') \in \symcal{K}_\nu}} P_{k'}P_{k''}P_{k'''} g(\phi_{k'} {+} \phi_{k''} {-} \phi_{k'''} {-} \phi_k).
	\end{align}
For large $M$, the terms for which the argument of the array gain $g(\cdot)$ is nonzero become   small, and it holds approximately that 
	\begin{align}
	\frac{R_{e_ke_k}[\ell]}{M^2} &\to |a_3|^2 \sum_{\mathclap{\nu=-1}}^{1}\! \gamma_{3,\nu}[\ell] P_k \sum_{k'=1}^{K} P_{k'}^2.
	\end{align}
    Because the direction of some of the distortion terms is parallel to the channel of the user, these terms do not vanish when the number of antennas is increased.  The application of other linear decoders (different from maximum-ratio combining) also does not help in this case.
	
	The autocorrelation will now be studied in a series of case studies, both with dominant in-band signals and out-of-band blockers, to illustrate how the distortion affects different system setups.  As a reference, the \textsc{lte} standard \cite[Tab.~7.1.1.1]{3GPP_TS36104_LTE_TxRx} requires a base stations to be able to handle interfering in-band signals that are approximately 55\,dB stronger than the desired signal and out-of-band 80\,dB stronger.
	
	\subsection{One User, One Blocker}
	In case there is only one served user, $K=1$, the index sets of the three frequencies $\nu=-1,0,1$ are
	\begin{align}
		\symcal{K}_{-1} &= \{(1,1,2)\}\\
		\symcal{K}_{0} &= \{(1,1,1), (2,1,2), (1,2,2)\}\\
		\symcal{K}_{1} &= \{(2,1,1), (1,2,1), (2,2,2)\},
	\end{align}
	and the autocorrelation of the error due to the distortion becomes:
	\begin{align}
		&R_{e_ke_k}[\ell] \notag\\
		&\quad= |a_3|^2 \Bigl( \gamma_{3,-1}[\ell] P_1^2 P_2 g(\phi_1 - \phi_2) + \gamma_{3,0}[\ell](P_1^3 + 2P_1P_2^2) M^2\\
		&\quad+ \gamma_{3,1}[\ell] (2P_1^2P_2 + P_2^3) g(\phi_2-\phi_1) \Bigr).\label{eq:01239391}
	\end{align}
	Depending on the relative powers between the user and the blocker and the number of antennas, only a few of these terms are significant.  In Table~\ref{tab:91092838109287}, four scenarios are specified: few or many antennas, weak or strong blocker.  
	It is interesting to note that in all cases there is at least one term that scales with the square of the number of antennas.
	These terms combine in the same way as the in-band signal because their spatial direction as given by \eqref{eq:9272322} is the the same as that of the in-band signal.  In other words, the error and the in-band signal will obtain the same array gain.  
	
	\begin{table}
		\ifdraft\linespread{1}\fi
		\centering
		
		\caption{Dominant terms in the autocorrelation $R_{e_1e_1}[\ell]$ in the presence of a blocker, whose received power is $P_2$, in a free-space line-of-sight scenario with one served user, whose received power is $P_1$.\label{tab:91092838109287}}

		\begin{tabular}{|c|c|c|}
			\hline
			& \parbox{8em}{\centering\ifdraft\setlength{\baselineskip}{.6\baselineskip}\fi few antennas\\ $M^2 \ll P_2/P_1$} & \parbox{8em}{\centering\ifdraft\setlength{\baselineskip}{.6\baselineskip}\fi\strut \smash{many} antennas\\ $M^2 \gg P_2/P_1$}\\
			\hline
			\parbox{5em}{\strut\centering\ifdraft\setlength{\baselineskip}{.6\baselineskip}\fi negligible blocker\\$P_1 \gg P_2$\strut} & $|a_3|^2 \gamma_{3,0}[\ell] P_1^3 M^2$ & $|a_3|^2 \gamma_{3,0}[\ell] P_1^3 M^2$\\
			\hline
			\parbox{5em}{\strut\centering\ifdraft\setlength{\baselineskip}{.6\baselineskip}\fi strong\\ blocker\\$P_1 \ll P_2$\strut} &\parbox{13em}{$2|a_3|^2 \gamma_{3,0}[\ell] P_1 P_2^2 M^2$\\\null\hfill$+ |a_3|^2 \gamma_{3,1}[\ell] P_2^3 g(\phi_2{-}\phi_1)$} & $|a_3|^2 \gamma_{3,0}[\ell] P_1P_2^2M^2$\\
			\hline
		\end{tabular}
	\end{table}
	
	\begin{observation}
		When the blocker is negligible, the autocorrelation is dominated by the $P_1^3$ term,
		\begin{align}
		R_{e_1e_1}[\ell] \approx |a_3|^2\gamma_{3,0}[\ell]P_1^3M^2,
		\end{align}
		which is temporally white.  This term grows with the number of antennas, $M$, at the same rate  as the linear signal.  The distortion therefore does not vanish 
		when $M$ is increased.
	\end{observation}
	
	\begin{observation}
		When the blocker is strong, the autocorrelation function of the error is approximately:
		\begin{align}
		R_{e_1e_1}[\ell] &\approx |a_3|^2 \Bigl( 2 \gamma_{3,0}[\ell]P_1P_2^2 M^2 + \gamma_{3,1}[\ell] P_2^3 g(\phi_2-\phi_1) \Bigr).\label{eq:9199386767}
		\end{align}
		The second term scales with $P_2^3$ and can possibly hurt the performance of the system significantly if $P_2$ is large.  
		The first term only scales with $P_2^2$ but it also combines in the same way as the desired signal, and scales with the number of antennas.  
		Hence, if the number of antennas is increased and the difference in sine angles between the blocker and the user is outside the narrow main lobe, $2\pi/M < |\phi_2-\phi_1| \mod 2\pi$, the relative attenuation $g(\phi_2 -\phi_1)/M^2$ goes to zero and the second term vanishes.  However, if the blocker stands such that its sine angle is inside the narrow main lobe of the user, $2\pi/M > |\phi_2-\phi_1| \mod 2\pi$, then $g(\phi_2-\phi_1) \propto M^2$ and the second term does not vanish.  In case of a strong blocker, the large number of antennas in massive \MIMO thus can alleviate the distortion by reducing the distortion power from being proportional to $P_2^3$ to being proportional to $P_2^2$.
	\end{observation}
	
	\subsection{Multiple Users, No Dominant User}
	A user $\chi$ is said to be \emph{dominant} if the received power $P_{\chi} \gg P_k$, for all other users $k\neq\chi$.  In this section, a system that performs power control is considered, such that there are no dominant users.  Furthermore, it is assumed that there is no blocker and that the number of antennas is large.  Then many of the gains $g(\cdot)$ in the sum \eqref{eq:91097811} can be assumed to be negligibly small for a user $k$ that has a unique angle of arrival, i.e.\ a $\phi_k \neq \phi_{k'}$ for $k'\neq k$.  Subsequently, only $2(K-1)$ terms combine coherently with maximum-ratio combining, i.e.\ have $g(\cdot) = M^2$, and the autocorrelation can be approximated as follows:
	\begin{align}
		R_{e_ke_k}[\ell] &= |a_3|^2 \gamma_{3,0}[\ell] \sum_{\mathclap{(k',k'',k''')\in\symcal{K}_0}} P_{k'} P_{k''} P_{k'''} g(\phi_{k'} + \phi_{k''} - \phi_{k'''} - \phi_{k})\\
		&\approx 2 |a_3|^2 \gamma_{3,0}[\ell] \Bigl(P_k^3 +  2 P_k \sum_{k'\neq k} P_{k'}^2\Bigr) M^2.
	\end{align}
	
	\begin{observation}
		The power of the error grows with the number of users, because the total received power grows with the number of users.  This stands in contrast to the amplifier-induced distortion in the downlink, where the coherent distortion scales inversely with the number of users \cite{mollen2017nonlinear2ArXiv}, because the power to each user decreases proportionally to $1/K$ assuming a fixed total radiated power.  
	\end{observation}
	
	\subsection{Multiple Users, One Dominant User}
	If there is no power control in the system, the served users might be received with widely different powers.  To illustrate this case, it will be assumed that one served user $\chi$ is dominant, $P_{\chi} \gg P_{k'}$ for all $k'\neq \chi$.  The significant terms in the autocorrelation of the distortion error are then the terms that contain the third power of the power of the dominant user $P_{\chi}^3$ and the terms with a large $g(\cdot)$ that contain the power $P_{\chi}$.  
	As in the single-user case, the autocorrelation of the dominant user is approximately
	\begin{align}
		R_{e_\chi e_{\chi}}[\ell] &\approx |a_3|^2 \gamma_{3,0}[\ell] P_{\chi}^3 M^2.
	\end{align}
	For the other, non-dominant users, $k\neq \chi$, the autocorrelation is
	\begin{align}
		R_{e_ke_k}[\ell] &\approx |a_3|^2 \gamma_{3,0}[\ell] \left(P_{\chi}^3 g(\phi_{\chi} - \phi_k) + 2 M^2 P_{k} P_{\chi}^2\right).
	\end{align}
	If the dominant user has a different incidence angle, $\phi_\chi \neq \phi_k$, using a large number of antennas thus removes the first term that scales with $P_\chi^3$.  The second term that scales with $P_\chi^2$ combines in the same way as the desired signal, however, and will not vanish with an increased number of antennas.  It is noted that placing a null in the direction of the dominant user would remove the term that scales with $P_\chi^3$ also with a finite number of antennas.
	
	\subsection{Multiple Users, One Blocker}
	If there is a blocker that is received with a much higher power than the served users, $P_{K+1}^2 \gg \sum_{k=1}^{K} P_k^2$, then the autocorrelation of the distortion error only contains a few significant terms, just as in the single-user case in \eqref{eq:9199386767}.  
	The terms containing the third power of the received power from the blocker $P_{K+1}^3$ and the terms with a large array gain that contain the second power $P_{K+1}^2$ are significant, and the autocorrelation is approximately:
	\begin{multline}
		R_{e_ke_k}[\ell] \\
		\approx |a_3|^2 \Bigl(2 \gamma_{3,0}[\ell] P_{K+1}^2 P_k M^2 + \gamma_{3,1}[\ell] P_{K+1}^3 g(\phi_{K+1} - \phi_k)\Bigr). \label{eq:8287221}
	\end{multline}
	If the user has an incidence angle that is different from the blocker, the second term that scales with $P^3_{K+1}$ becomes negligible when the number of antennas is increased.  
	The first term that scales with $P_{K+1}^2$, however, remains as it scales with $M^2$.
	
	If the blocker stands inside the main lobe of the served user $k$ (i.e.\ $2\pi/M > |\phi_{K+1} - \phi_k| \mod 2\pi$), both terms in \eqref{eq:8287221} will scale with $M^2$ and the autocorrelation is:
	\begin{align}
		R_{e_ke_k}[\ell] &\approx |a_3|^2 M^2 \Bigl( 2 \gamma_{3,0}[\ell] P_{K+1}^2 P_k + \gamma_{3,1}[\ell] P_{K+1}^3 \Bigr).
	\end{align}
	The second term that has a temporal correlation that is colored is the larger of the two terms if $P_{K+1}>2P_k\gamma_{3,0}[0]/\gamma_{3,1}[\ell]$.

	\section{Different Amplifiers}
	Due to fabrication imperfections, the amplifiers might not all be equal.  How linearity variations among amplifiers affect the spatial pattern of the distortion is studied in this section.  We note that, since the channel is estimated through the same \LNAs, the effects of the first-degree coefficients $\{a_{1m}\}$ are adjusted for in the decoding and the spatial patterns of the desired signals are not significantly affected by hardware variations.
	
	The variations between the amplifiers are modeled as independent random deviations $\alpha_m \sim \CN(0,\eta |a_3|^2)$ from a common mean $\sqrt{1-\eta} a_3$:
	\begin{align}
		a_{3m} a^*_{1m} = \sqrt{1-\eta} a_3 + \alpha_m,\label{eq:395728}
	\end{align}
	where the parameter $0 \leq \eta \leq 1$ describes the degree of deviation between amplifiers.  This model should be seen as a first-order approximation, short of deviation models based on measurements on a batch of \LNAs.  Other models, such as modeling the deviations as random phase shifts of a common mean, give similar results (not shown here due to space limitations).
	
	If maximum-ratio combining is employed and no attempt is made to make the error due to distortion \eqref{eq:0091929} small, the weights
	\begin{align}
	w_{km} = a^*_{1m} h^*_{km}
	\end{align}
	are used.  Then the random array gain of the distortion becomes:
	\begin{align}
	G(\varphi) &\defas \left|\sum_{m=1}^{M} a_{3m} a^*_{1m} e^{jm\varphi}\right|^2\\
	&= \left|\sum_{m=1}^{M} \sqrt{1-\eta} a_3 e^{jm\varphi} + \sum_{m=1}^{M} \alpha_m e^{jm\varphi}\right|^2.
	\end{align}
	The expectation of this random array gain is:
	\begin{align}
		\Exp[G(\varphi)] &= (1-\eta) |a_3|^2 g(\varphi) + M\eta|a_3|^2\\
		&=|a_3|^2 \left(g(\varphi) (1-\eta) + M\eta\right).\label{eq:76831}
	\end{align}
	
	The expectation in \eqref{eq:76831} is illustrated in Figure~\ref{fig:array_gain_different_LNAs}\,a, which shows the envelope $\psi(\varphi)$ of $g(\varphi)$.  It can be seen that distortion at angles for which $\varphi \neq 0$ is increasingly picked up by the spatial filter as the degree of deviation $\eta$ is increased.  The average array gain of the coherent terms, for which $\varphi=0$, is given by
	\begin{align}
	\Exp\left[G(0)\right] &= |a_3|^2 M^2 (1 -\eta(1 - 1/M)).
	\end{align}
	This expectation is shown in Figure~\ref{fig:array_gain_different_LNAs}\,b.  
	It can be seen how the array gain of the distortion that combines coherently with maximum-ratio combining decreases with an increasing degree of deviation $\eta$ among the amplifiers.  However, the degrees of deviation $\eta$ has to be fairly large in order to observe a significant reduction in the array gain of the coherent distortion.
	
	\begin{figure}
		\centering
		\ifdraft
		\setlength{\figbredd}{.5\linewidth}
		\pgfplotsset{height=34ex,width=\figbredd}
		\else
		\setlength{\figbredd}{\linewidth}
		\fi
		\begin{minipage}{\figbredd}%
		\centering%
\\[\abovecaptionskip]\footnotesize
		a) Envelope of the mean array gain of the distortion\ifdraft\else\bigskip\fi
		\end{minipage}\ifdraft\else\newline\pgfplotsset{height=\flatFiguresHeight}\fi%
		\begin{minipage}{\figbredd}%
			\centering%
			%
\\[\abovecaptionskip]\footnotesize
			b) Average array gain of the coherent distortion
		\end{minipage}
		\ifdraft\vspace{1ex}\fi
		\caption{The mean array gain of the distortion in arrays with variable \LNAs and $M = 100$ antennas.  Around $\varphi = 0$ the gain is not visible from the envelope in Figure~a, instead it is shown in Figure~b.  The envelope for $\eta=0.1$ can be compared to the actual mean array gain, which is shown by the grey curve in Figure~a.}
		\label{fig:array_gain_different_LNAs}
	\end{figure}
		
	The result of variations in the amplifier linearity is that the array gain of the coherent distortion is decreased and that the noncoherent distortion is not suppressed as much.  
	
\section{Comparison with Spatially-Uncorrelated Distortion Model}

It is instructive to compare our findings with those that result from
other models.  Specifically, a range of previous work has modeled the signal distortion from non-ideal
hardware in \MIMO systems as spatially uncorrelated additive noise
in discrete, symbol-sampled
time \cite{bjornson2013massive,schenk2008rf,bjornson2017massiveBook}.  The spatially uncorrelated distortion model is
analytically attractive as it enables the derivation of
closed-form performance expressions.  It is, furthermore, known that the in-band error-vector
magnitude predicted by the spatially uncorrelated distortion model matches that of
simulations with a behavioral amplifier model in cases where many
users are concurrently multiplexed over a Rayleigh fading channel \cite{UGUSGC14}.   

Yet, interestingly, the spatially uncorrelated distortion model also has some implications that
contradict the findings of our work. For example, Corollary~4
in \cite{bjornson2015massive} implies that one can tolerate stronger
hardware imperfections as the number of antennas $M$ increases and, hence, that one can relax the linearity constraints and hardware quality as $M$ increases. In contrast, our analysis   shows that parts of the distortion can obtain an array gain of $M^2$,
pointing to the opposite effect: hardware quality cannot be relaxed arbitrarily much as $M$ is increased. 

In what follows, we expound on this discord in some more detail and compare our conclusions with the spatially uncorrelated distortion model. 
While doing that, two caveats must be kept in mind.  
First, the spatially uncorrelated distortion model is, to our
understanding, not intended to give an accurate behavioral description
of an \LNA specifically, although, partially, the purpose of the model is to describe the effects of
nonlinearities in the transceiver chain.  
Second, while hardware
distortion certainly is a property of the hardware and not of the
propagation channel, to our knowledge, the spatially uncorrelated distortion model has been applied
only to the analysis of in-band distortion in
Rayleigh fading channels in previous work---whereas much of our
analysis focuses on out-of-band signals and more general channel models.

Let us consider a fixed channel response, ${h_{km}}$, or a fading channel  conditioned on one of the channel states.
According to the spatially uncorrelated distortion model, the distortion is a zero-mean
Gaussian random variable with the following correlation function (see, e.g., \cite[Eqs.~(6) and (8)]{bjornson2013massive}):
	\begin{align}\label{eq:neweq1}
		R_{d_md_{m'}}[\ell] = \begin{cases}
		0, & m\neq m' \text{ or }\ell \neq 0\\
		\kappa \sum_{k=1}^{K} |h_{km}|^2 P_k, &\text{otherwise},
		\end{cases}
	\end{align}
	where $\kappa$ is a parameter that models the quality of the hardware. (A small value of $\kappa$ represents accurate and expensive hardware.)
In contrast, starting from our modeling framework,   ignoring the temporal correlation
given by $\{\gamma_{3,\nu}[\ell]\}$ in \eqref{eq:90019}, which the spatially uncorrelated distortion model neglects, and  considering only the significant terms for which $(\ell,\nu) = (0,0)$,
the spatial correlation of the third-degree distortion term is:
	\begin{align} 
		R_{u_{3m}u_{3m'}}[0] = a_{3m} a^*_{3m'} \gamma_{3,0}[0] \sum_{\mathclap{k,k',k''\in\symcal{K}_0}} P_kP_{k'}P_{k''} \bar{h}_{kk'k''m} \bar{h}^*_{kk'k''m'}.\label{eq:83927}
	\end{align}
A comparison between (\ref{eq:neweq1}) and (\ref{eq:83927}) reveals two important  qualitative differences:
(i) According to the spatially uncorrelated distortion model, \eqref{eq:neweq1}, the distortion is uncorrelated among the antennas, but in \eqref{eq:83927}, the cross-correlation is
a function of the channels (via $\{\bar{h}_{kk'k''m}\}$).
As shown earlier for maximum-ratio combining, 
the distortion obtains an array gain with our model, whereas it does not with the spatially uncorrelated distortion model, because
the distortion in \eqref{eq:neweq1} is spatially white. 
(ii) According to the spatially uncorrelated distortion model, the distortion power scales linearly with the received power, but in \eqref{eq:83927}, the power scales non-linearly;
since $a_{3m}$ is power dependent, the power scales with $|a_{3m}|^2 \sum P_k P_{k'} P_{k''} |\bar{h}_{kk'k''m}|^2$.  
These observations hold conditioned on ${h_{km}}$, hence independently of the propagation channel model.

To understand the difference more intuitively in terms of the spatial pattern of the array, consider the special case of a free-space line-of-sight channel. Denote by
	\begin{align}
		e_k = \sum_{m=1}^{M} d_m w_{mk},
	\end{align}
	 the error due to distortion, where
         the decoding weights $\{w_{mk}\}$ are functions of the channel response, given by  \eqref{eq:907761788992}.
The error under the spatially uncorrelated distortion model has the variance
	\begin{align}
		\Exp\left[|e_k|^2 \Bigm| \{h_{km}\}\right] &= \kappa \sum_{k'=1}^{K} P_{k'} \sum_{m=1}^{M} |h_{k'm}|^2 |w_{mk}|^2\\
        &= \kappa M \sum_{k'=1}^{K}P_{k'} .\label{eq:83397711}
	\end{align}
The interpretation is that the array gain of the distortion is $\Exp[G(\varphi)]/\kappa = M$ independently of the normalized sine angles $\{\phi_k\}$,
see Figure~\ref{fig:array_gain_different_LNAs}. The spatial pattern resulting from the spatially uncorrelated distortion assumption is thus constant and
does not show the array gain of the distortion terms with the same spatial characteristics as the desired signal, nor does it show the suppression of the distortion terms with different spatial characteristics.  

	\section{Frequency-Selective Channels}
	When the fading is frequency-selective and isotropic, the spatial pattern is harder to illustrate than in the free-space line-of-sight case, where a spatial pattern like the one in Figure~\ref{fig:approximate_sum} gives a good picture.  In this section, we   numerically study the effects of a blocker on a single-user system with frequency-selective fading for different number of receiving antennas.
	
	To simulate a frequency-selective propagation environment, it is assumed that the transmit signal from transmitter $k$ is received from $V$ scattering clusters, each located at the position $(\symfrak{x}_{kv}, \symfrak{y}_{kv})$.  Then, instead of the expression in \eqref{eq:88181833}, the received signal is given by
	\begin{align}
		u_m(t) = \sum_{k=1}^{K+1}\sum_{v=1}^{V} h_{kv} e^{-j2\pi(\tau_{kv}+\epsilon_{kv})} x_k(t-\tau_{kv}-\epsilon_{kv}) s_m(\symfrak{x}_{kv},\symfrak{y}_{kv}).
	\end{align}
	As the aim is to study the spatial characteristics of the distortion, the thermal noise is assumed to be zero.  The path losses $\{h_{kv}\}$ and delays $\{\tau_{kv}\}$ are assumed to be given by the environment and fixed.  The small variations in the delays between coherence intervals are assumed to cause a random phase shift that is uniformly distributed:
	\begin{align}
		\epsilon_{kv} \sim \operatorname{uniform}[0,2\pi].
	\end{align}
	It will be assumed that the array is located along the $\symfrak{y}$ axis and is a uniform linear array with an antenna spacing $\Delta = \lambda/2$ that equals half the wavelength of the carrier frequency.  Under this assumption, element $m$ of the steering vector is given by:
	\begin{align}
		s_m(\symfrak{x}, \symfrak{y}) \defas e^{-j2\pi (\sqrt{\symfrak{x}^2 + \symfrak{y}^2} - \sqrt{\symfrak{x}^2 + (\symfrak{y} - (m-1) \Delta)^2}) / c},
	\end{align}
	where $c$ is the speed of light.  
	
	The number of paths was chosen to $V=10$ and the channel parameters were taken as one realization of the random variables
	\begin{align}
		\tau_{kv} &\sim \operatorname{uniform}[0,\sigma_\tau]\\
		h_{kv} &\sim \operatorname{Rayleigh}(1/V)\label{eq:929283}\\
		\symfrak{x}_{kv} &\sim \operatorname{uniform}[100\lambda,5000\lambda]\\
		\symfrak{y}_{kv} &\sim \operatorname{uniform}[-5000\lambda,5000\lambda],
	\end{align}
	where the delay spread was chosen to be $\sigma_\tau = \SI{3}{\micro\second}$, which corresponds to a difference in path lengths of \SI{900}{m}. The carrier frequency was assumed to be $f_\text{c} = \SI{2}{GHz}$ and thus the wavelength $\lambda = \SI{.15}{m}$.  
	
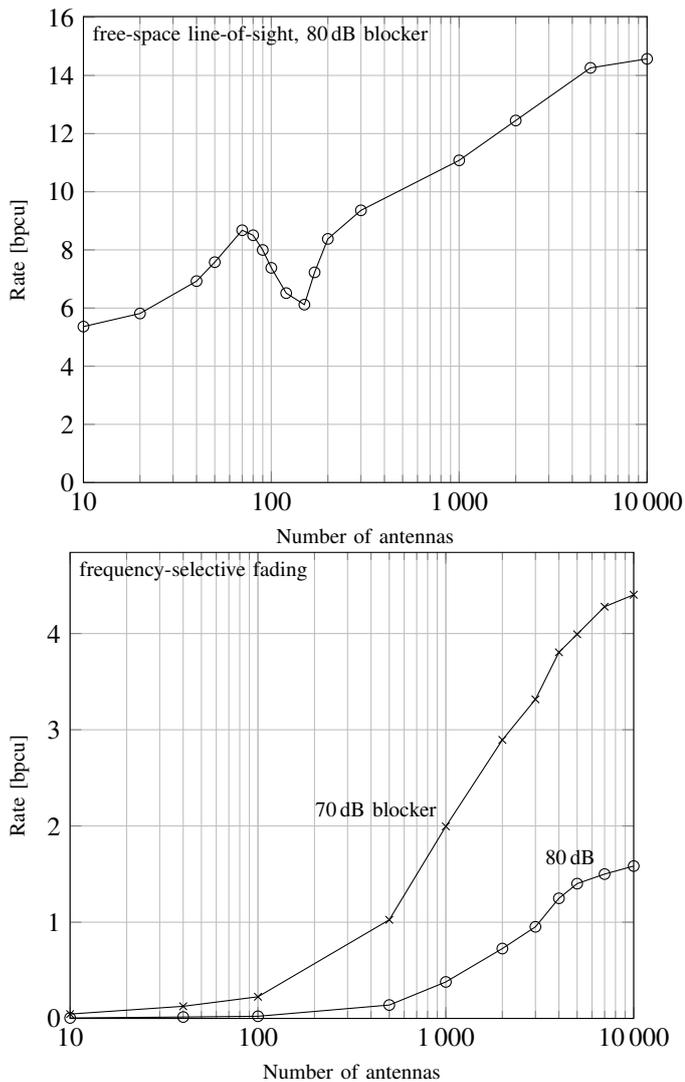
\begin{figure}
		\ifdraft\pgfplotsset{height=24ex,width=.49\linewidth}\hspace{-2em}\fi
		\centering
		\begin{tikzpicture}
		\begin{semilogxaxis}[
		xlabel = {Number of antennas},
		ylabel = {Rate [bpcu]},
		xmin = 10,
		xmax = 10000,
		ymin = 0,
		xtick = {10, 100, 1000, 10000},
		scaled ticks=false,
		log ticks with fixed point,
		x tick label style={/pgf/number format/1000 sep=\,},
		grid = both,
		]
		\addplot[
		mark = o,
		color=black,
		]
		table[x=nr_antennae, y=rate, row sep=crcr]{
			nr_antennae rate\\
			10000 14.57\\
			5000 14.26\\
			2000 12.45\\
			1000 11.08\\
			300 9.36052992667\\
			200 8.37875357522\\
			170 7.22744903541\\ 
			150 6.11676974704\\ 
			120 6.51368518222\\ 
			100 7.38226415157\\
			90 7.99748366269\\
			80 8.50361521938\\
			70 8.6758503882\\ 
			50 7.57747200043\\ 
			40 6.92518961311\\
			20 5.81276715655\\
			10 5.36239740547\\
		};
		\node[anchor=north west, font = \footnotesize] at (rel axis cs: 0.0, 1.0) {free-space line-of-sight, 80\,dB blocker};
		\end{semilogxaxis}
		\end{tikzpicture}\ifdraft\else\newline\fi%
		\begin{tikzpicture}
			\begin{semilogxaxis}[
			xlabel = {Number of antennas},
			ylabel = {Rate [bpcu]},
			xmin = 10,
			xmax = 10000,
			ymin = 0,
			xtick = {10, 100, 1000, 10000},
			scaled ticks=false,
			log ticks with fixed point,
			x tick label style={/pgf/number format/1000 sep=\,},
			tick label style={/pgf/number format/fixed},
			grid = both,
			]
			\addplot[
			mark = x,
			color=black,
			]
			 table[row sep=crcr]{
				10000 4.40290890832\\
				7000 4.27847659129\\
				5000 3.99272305329\\
				4000 3.80508478901\\
				3000 3.31615153528\\
				2000 2.89492755027\\
				1000 1.9957\\
				500 1.02492424669\\
				100 0.224904683168\\
				40 0.125852428268\\ 
				10 0.0454571311016\\ 
				};
			\node[anchor=south east, font = \footnotesize] at (axis cs: 1000, 2) {70\,dB blocker};
			\addplot[
			mark = o,
			color=black,
			]
			table[x=nr_antennae, y=rate, row sep=crcr]{
				nr_antennae rate\\
				10000 1.58344664904\\
				7000 1.50010440196\\
				5000 1.40081398988\\
				4000 1.24839612364\\
				3000 0.952036600065\\
				2000 0.726178487502\\
				1000 0.379082252828\\
				500 0.13903217523\\
				100 0.0222553560827\\
				40 0.0146058576142\\ 
				10 0.00482918115026\\ 
			};
			\node[anchor=south east, font = \footnotesize] at (axis cs: 7000, 1.5) {80\,dB};
			\node[anchor=north west, font = \footnotesize] at (rel axis cs: 0.0, 1.0) {frequency-selective fading};
			\end{semilogxaxis}
		\end{tikzpicture}
		\caption{The rate in \eqref{eq:897376688} in a system with one served user and a blocker that is received with a higher power than the served user.  The amplifiers are run \SI{8}{dB} below the one-dB compression point on average.  The coefficients from Table~\ref{tab:amp_coeffs} are used.}
		\label{fig:saturating_rate}
	\end{figure}

	The effect of a blocker on the performance given by the rate in \eqref{eq:897376688} is shown in Figure~\ref{fig:saturating_rate}.  It can be seen that the performance saturates as the number of antennas is increased.  This is due to the coherent distortion from the blocker that is present also in this frequency-selective setting.  
	
	The result can be compared to the results for a free-space line-of-sight ($V=1$) setting in Figure~\ref{fig:saturating_rate}.  It can be seen that the rate saturates at a much higher level in the line-of-sight case, since in this case the distortion from the blocker is concentrated to the four lobes in \eqref{eq:01239391}.  
	The precise effect of these lobes depends on the array geometry and changes with the number of antennas, which explains the behavior of the rate around $M=100$.
	In the frequency-selective case, however, the distortion is spread in many more directions and is harder to suppress using a spatial filter.  Furthermore, the frequency-selective fading can cause individual amplifiers to operate much closer to saturation than the average amplifier, which causes excessive distortion.
	
	\section{Conclusion}
We have presented a framework for rigorous analysis of the effects of   amplifier nonlinearity distortion in the massive \MIMO uplink,
and specifically for the analysis of the effects of a strong blocker.  
The analysis is based on orthogonal polynomials.  For simplicity, we have assumed Gaussian signals in order to use the well-documented Itô-Hermite polynomials, leaving the possible
study of other signals for future work.  
The main conclusions are as follows. 

	The impact of a blocker is qualitatively, and quantitatively, different in the \MIMO case as compared to in a single-antenna system.  Specifically,  in a massive \MIMO base station with a large number of antennas,
	the effective power of the distortion created by a blocker scales quadratically rather than cubically with the power of the blocker,
	assuming spatial matched filtering (maximum-ratio combining).  

The   distortion resulting from nonlinearities is spatially correlated, and the effects of this correlation must be adequately taken into account
in order to obtain accurate conclusions. Specifically, models that assume spatially uncorrelated distortion can yield
incorrect conclusions in some important cases of interest when the number of antennas is large, especially in free-space line-of-sight propagation.   
		
In the case of a single served user in free-space line-of-sight without blockers (and assuming that all amplifiers are identical and the received power at each \LNA is the same), any decoder, for which the signal of interest combines coherently, will make the distortion combine fully coherently and the effect of \LNA nonlinearities is the same as in a \SISO system without spatial processing.  
In this free-space line-of-sight case, the signal-to-distortion ratio cannot be improved by linear processing alone.
This limits the effective \SINR that can be achieved, even if the number of antennas is increased.  
In other 
 propagation scenarios, however, the impact of distortion can be reduced by spatial filtering, rendering the  effects of the distortion  
  smaller than in a single-antenna system. 

While our study considered spatial matched filtering, other types of linear
 processing may mitigate the distortion to a greater
extent---except for the single-user free-space line-of-sight case, in which this is impossible as already pointed out. 

We hope that the rigorous continuous-time models presented here will
stimulate the development of improved signal processing algorithms to deal with the important task of blocker suppression in 
massive \MIMO systems with nonlinear receiver frontends.  
Practical issues such as channel estimation,
 and characterization of the nonlinear transfer characteristics of the frontend, will require
  attention in order for such processing to properly work.  
	
\ifdraft\vspace{-3ex}\fi
	
	\appendices
	\section{Proof of Theorem~\ref{the:891856178960781}}\label{app:29038}
First the gain in \eqref{eq:1992811123331} is investigated with $w_{km} = a^*_{1m} h^*_{km}$:
\begin{align}
g_k &= \sum_{m=1}^{M} |a_{1m}|^2 \Exp\left[|h_{km}|^2\right]\\
&= \sum_{m=1}^{M} |a_{1m}|^2.
\end{align}
The square of this is $|g_k|^2 = G_k \bigl|\sum_{m=1}^{M}|a_{1m}|^2\bigr|^2/M^2$.

Then the interference variance in \eqref{eq:19999911} is considered:
\begin{align}
\tilde{I}_{kk'} &= \operatorname{var}\left(\sum_{m=1}^{M} |a_{1m}|^2 h^*_{km} h_{k'm}\right)\\
&= \sum_{m=1}^{M} |a_{1m}|^4 \operatorname{var}\left(  h^*_{km} h_{k'm}\right)\\
&= \bar{I}_{kk'} \frac{\sum_{m=1}^{M} |a_{1m}|^2}{M},
\end{align}
because the channels are independent across $m$.  Because the channel coefficients are identically distributed across $m$, the variances $\operatorname{var}\left( h^*_{km} h_{k'm} \right)$ are equal for all $m$.  The variance of the nonlinear system is therefore equal to the variance in \eqref{eq:19988281822} of the linear system: $\tilde{I}_{kk'} = \bar{I}_{kk'}$.  

Similarly, the noise term has the variance:
\begin{align}
\operatorname{var}\left(\sum_{m=1}^{M} |a_{1m}|^2 w_{km} z_m[n]\right) &= \sum_{m=1}^{M} |a_{1m}|^4 \underbrace{\operatorname{var}\left( w_{km}z_m[n]\right)}_{=N_0/T}\\
&= \frac{N_0}{T} \sum_{m=1}^{M} |a_{1m}|^4.
\end{align}
By normalizing this with respect to the total amplification $\sum_{m=1}^{M}|a_{1m}|^2$ the noise variance is obtained and the theorem follows.

\ifdraft\vspace{-3ex}\enlargethispage*{\baselineskip}\fi
	
	\section{Derivation of the Polynomial Model}\label{app:polynomial_model}
The polynomial model in \eqref{eq:polynomial_model} can be derived from the memoryless polynomial passband description in \eqref{eq:real_pb_nonlinearity} by writing the passband signal in terms of its equivalent baseband signal:
\begin{align}
\hat{x}(t) = x(t) e^{j2\pi f_\text{c} t} + x^*(t) e^{-j2\pi f_\text{c} t}.
\end{align}
The $\varpi$\mbox{-}th power of the input signal is then:
\begin{align}
\hat{x}^\varpi(t) = \sum_{n=0}^{\varpi} {\varpi \choose n} x^n(t) (x^*(t))^{\varpi-n} e^{j2\pi f_\text{c} (2n-\varpi)t}.
\end{align}
If we apply the same lowpass filter as in \eqref{eq:baseband_signal} to this signal, it will filter out all terms except the term when $2n - \varpi = 1$.  Thus, $\symcal{B}(\hat{x}^\varpi(t) e^{-j2\pi f_\text{c}t}) = 0$ for all even $\varpi$, and:
\begin{align}
	\symcal{B}\left(\hat{x}^\varpi(t) e^{-j2\pi f_\text{c}t}\right) = {\varpi \choose \frac{\varpi+1}{2}} x(t) |x(t)|^{\varpi-1},
\end{align}
for odd $\varpi$, because the bandwidth of $x(t) |x(t)|^{\varpi-1}$ is smaller than the cutoff frequency of the filter $\Pi B/2 < f_\text{c}/2$.
	
	\section*{Source Code}
	The code used in the generation of some of the figures can be found at \url{https://github.com/OOBRadMIMO/LNAMaMIMO}.
	
	\ifCLASSOPTIONcaptionsoff
	\newpage
	\fi
	
	\bibliographystyle{IEEEtran}
	\ifdraft\enlargethispage*{\baselineskip}\vspace{-2ex}\fi
	\bibliography{bib_forkort_namn,bibliografi}
	\ifdraft\enlargethispage*{\baselineskip}\fi
	
	\begin{IEEEbiography}[{\includegraphics[width=1in,height=1.25in,clip,keepaspectratio]{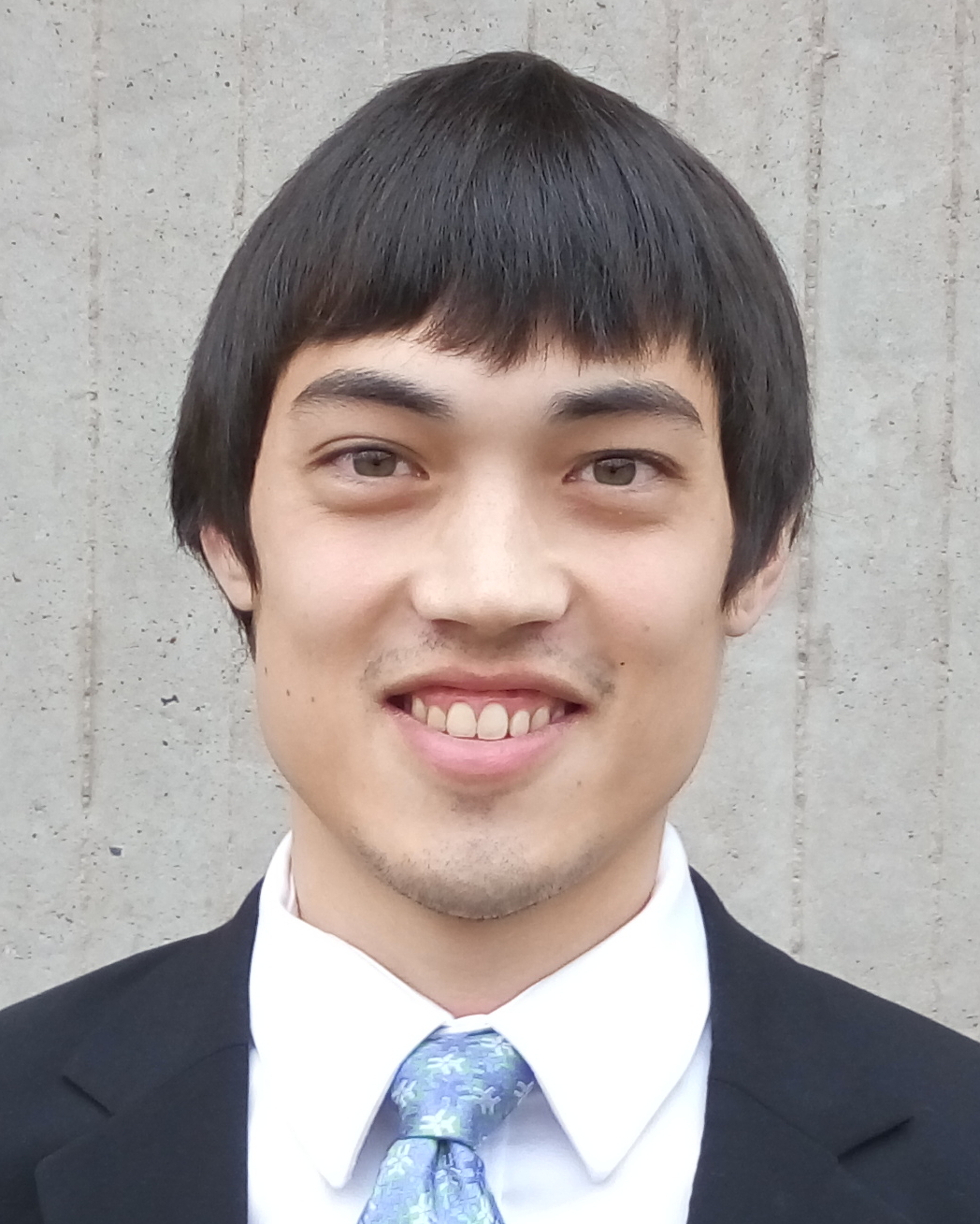}}]{Christopher Mollén}
	received the Ph.D. degree (2018) and the M.Sc. degree (2013) from Linköping University, Sweden.  His Ph.D.\ thesis \emph{High-End Performance with Low-End Hardware: Analysis of Massive MIMO Base Stations Transceivers} explored low-complexity hardware implementations of massive \MIMO base stations, including low-\PAR precoding, low-resolution \ADCs, and nonlinear amplifiers.  Previously he has worked as intern at Ericsson in Kista, Sweden, and in Shanghai, China.  From 2011 to 2012, he studied at the Eidgenössische Technische Hochschule (ETH) Zürich, Switzerland, as an exchange student in the Erasmus Programme.  And from 2015 to 2016, he visited the University of Texas at Austin as a Fulbright Scholar.  This work was done while at Linköping University.  He is now with Apple, 3D Vision.
	\end{IEEEbiography}
	
	\begin{IEEEbiography}[{\includegraphics[width=1in,height=1.25in,clip,keepaspectratio]{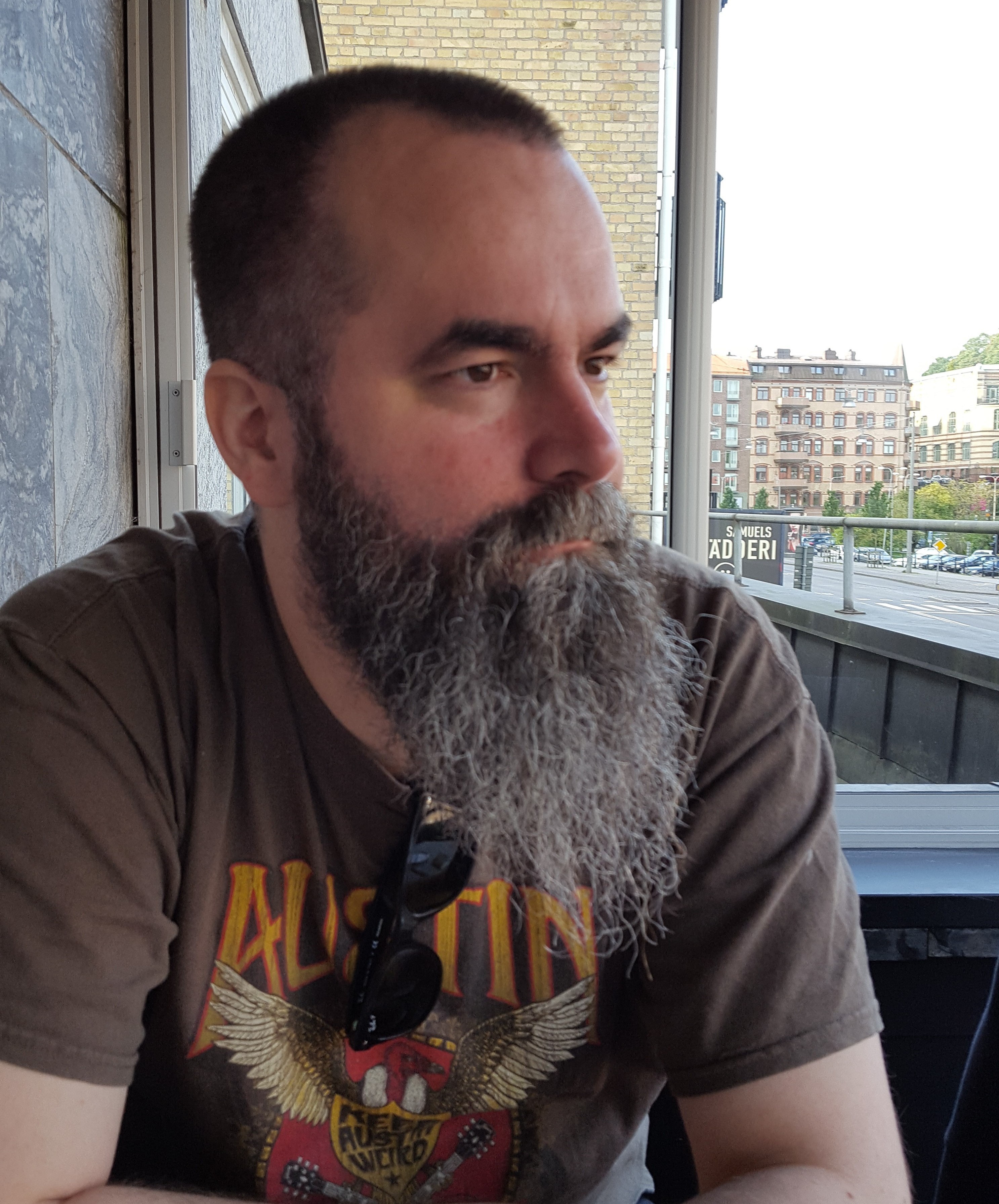}}]{Ulf Gustavsson}
		received the M.Sc.\ degree in electrical engineering from Örebro University, Örebro, Sweden, in 2006, and the Ph.D.\ degree from the Chalmers University of Technology, Gothenburg, Sweden, in 2011. He is currently a Senior Specialist with Ericsson Research where his research interests include radio signal processing techniques for hardware impairment mitigation and behavioral modeling of radio hardware for future advanced antenna systems. Dr.\ Gustavsson is currently also the lead scientist from Ericsson Research in the Marie Skłodowska-Curie European Industrial Doctorate Innovative Training Network, SILIKA (\url{http://silika-project.eu/}).
	\end{IEEEbiography}
	
	\begin{IEEEbiography}[{\includegraphics[width=1in,height=1.25in,clip,keepaspectratio]{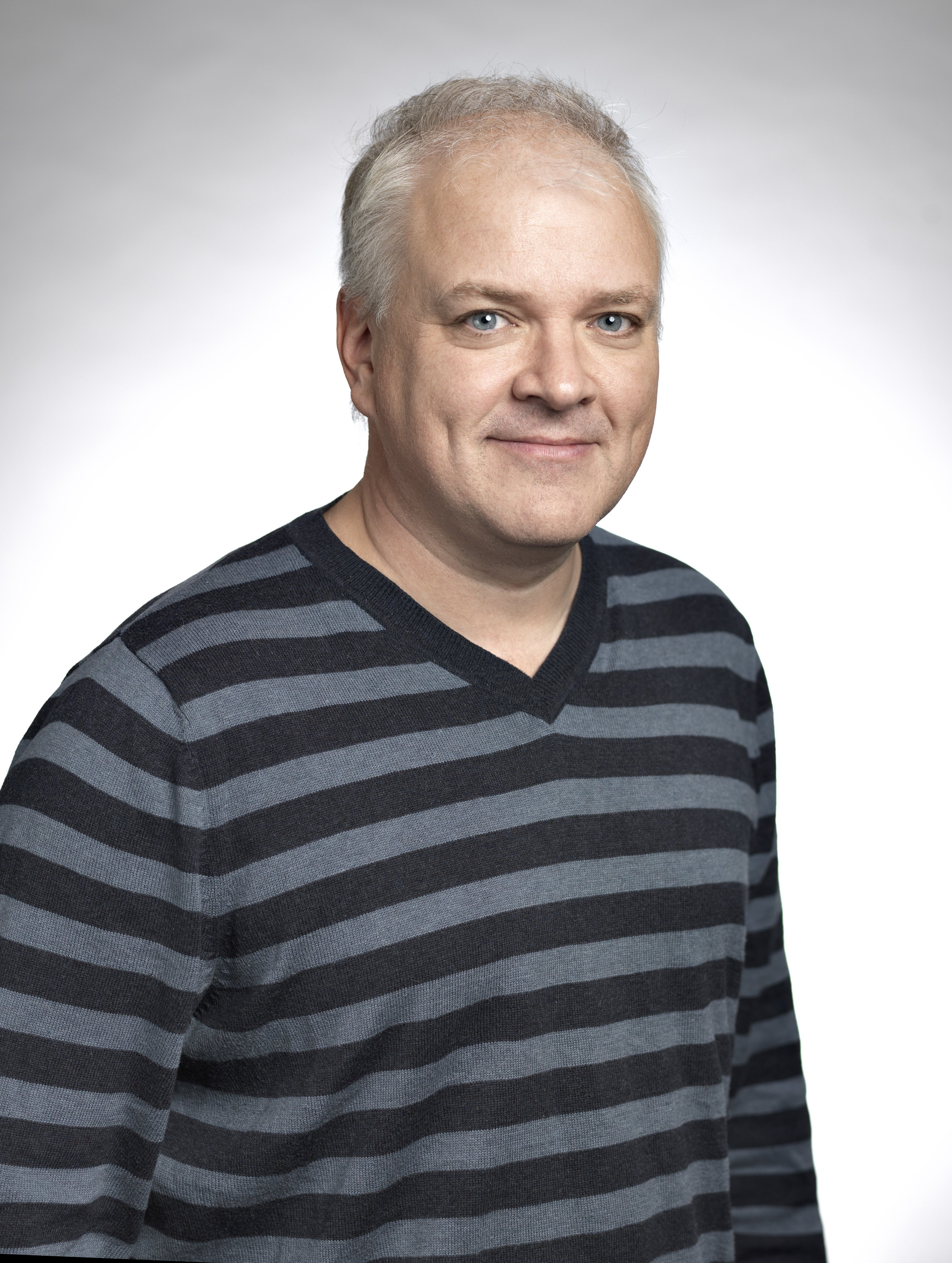}}]{Thomas Eriksson}
		received the Ph.D.\ degree in Information Theory from Chalmers University of Technology, Gothenburg, Sweden, in 1996. From 1990 to 1996, he was at Chalmers. In 1997 and 1998, he was at AT\&T Labs - Research, Murray Hill, NJ, USA. In 1998 and 1999, he was at Ericsson Radio Systems AB, Kista, Sweden. Since 1999, he has been with Chalmers University, where he is currently a professor of communication systems. Further, he was a guest professor with Yonsei University, S. Korea, in 2003-2004. He has authored or co-authored more than 200 journal and conference papers, and holds 12 patents.

		Prof.\ Eriksson is leading the research on hardware-constrained communications with Chalmers University of Technology. His research interests include communication, data compression, and modeling and compensation of non-ideal hardware components (e.g. amplifiers, oscillators, and modulators in communication transmitters and receivers, including massive \MIMO). Currently, he is leading several projects on e.g. 1) massive \MIMO communications with imperfect hardware, 2) \MIMO communication taken to its limits: 100Gbit/s link demonstration, 3) mm-Wave \MIMO testbed design, 4) Satellite communication with phase noise limitations, 5) Efficient and linear transceivers, etc. 

		He is currently the Vice Head of the Department of Signals and Systems with Chalmers University of Technology, where he is responsible for undergraduate and master's education. 
	\end{IEEEbiography}
	
	\begin{IEEEbiography}[{\includegraphics[width=1in,height=1.25in,clip,keepaspectratio]{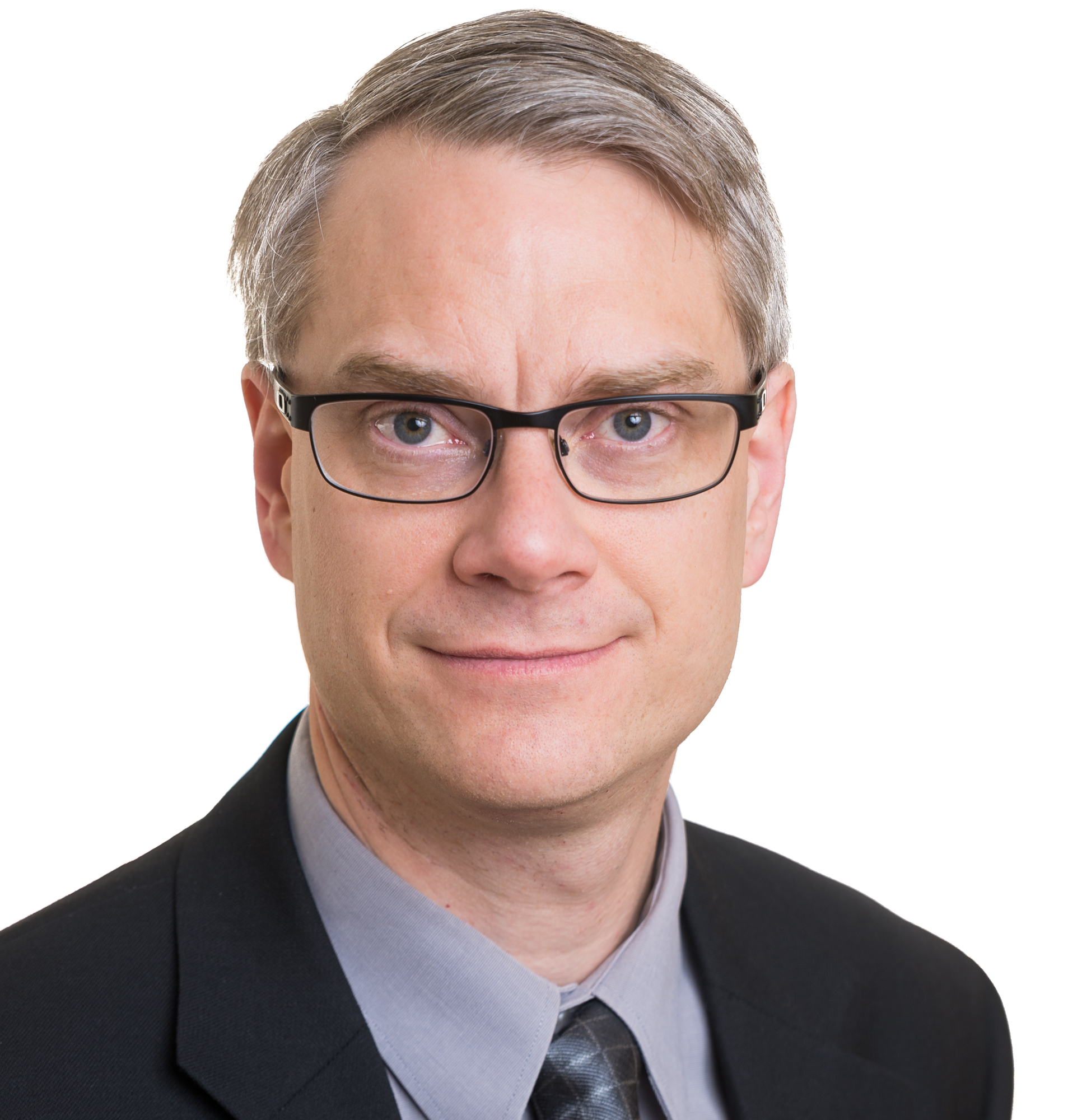}}]{Erik G. Larsson (S'99--M'03--SM'10--F'16)}
		received the Ph.D.\ degree from Uppsala University,
		Uppsala, Sweden, in 2002.
		
		He is currently Professor of Communication Systems at Link\"oping
		University (LiU) in Link\"oping, Sweden. He was with the KTH Royal
		Institute of Technology in Stockholm, Sweden, the George Washington
		University, USA, the University of Florida, USA, and Ericsson
		Research, Sweden.  In 2015 he was a Visiting Fellow at Princeton
		University, USA, for four months.  His main professional interests are
		within the areas of wireless communications and signal processing. He
		has co-authored some 150 journal papers on these topics, he is
		co-author of the two Cambridge University Press textbooks
		\emph{Space-Time Block Coding for Wireless Communications} (2003) and
		\emph{Fundamentals of Massive MIMO} (2016). He is co-inventor on 18
		issued and many pending patents on wireless technology.
		
		He is a member of the IEEE Signal Processing Society Awards Board
		during 2017--2019.  He is an editorial board member of the
		\emph{IEEE Signal Processing Magazine} during 2018--2020. 
		From 2015 to 2016 he served as chair of the IEEE
		Signal Processing Society SPCOM technical committee.  From 2014 to
		2015 he was chair of the steering committee for the \emph{IEEE
			Wireless Communications Letters}.  He was the General Chair of the
		Asilomar Conference on Signals, Systems and Computers in 2015, and its
		Technical Chair in 2012.  He was Associate Editor for, among others,
		the \emph{IEEE Transactions on Communications} (2010-2014) and the
		\emph{IEEE Transactions on Signal Processing} (2006-2010).
		
		He received the IEEE Signal Processing Magazine Best Column Award
		twice, in 2012 and 2014, the IEEE ComSoc Stephen O. Rice Prize in
		Communications Theory in 2015, the IEEE ComSoc Leonard G. Abraham
		Prize in 2017, and the IEEE ComSoc Best Tutorial Paper Award in 2018. 
		
		He is a Fellow of the IEEE.
	\end{IEEEbiography}
\end{document}